\documentclass{aa}
\usepackage{gensymb}

\usepackage{graphicx}
\usepackage{txfonts}

\newcommand{\gcc}{\mbox{ g cm$^{-3}$}}

\usepackage{subcaption}
\usepackage[colorlinks=true,linkcolor=blue,citecolor=blue,breaklinks=true]{hyperref}

\begin{document}

  \renewcommand{\vec}[1]{\mathbf{#1}}

   \title{Dust evolution during a protostellar collapse: Influence on  the coupling between the neutral gas and  magnetic field} 
  
   \author{ V. Vallucci-Goy,
          \inst{1}
          U. Lebreuilly,
          \inst{1}
          P. Hennebelle,
          \inst{1}
 }

   \institute{Université Paris-Saclay, Université Paris Cité, CEA, CNRS, AIM, 91191, Gif-sur-Yvette, France 
              \email{valentin.goy@cea.fr}      
             }

   \date{}

% \abstract{}{}{}{}{} 
% 5 {} token are mandatory
 
  \abstract
  % context 
   {The extent of the coupling between the magnetic field and the gas during the collapsing phase of star-forming cores is strongly affected by the dust size distribution, which is expected to evolve by means of coagulation, fragmentation, and other collision outcomes.}
  % aims. 
   {We aim to investigate the influence of key parameters on the evolution of the dust distribution, as well as on the magnetic resistivities during protostellar collapse.}
  % methods. 
   {We performed a set of collapsing single-zone simulations with \texttt{shark}. The code computes the evolution of the dust distribution, accounting for different grain growth and destruction processes, with the grain collisions being driven by brownian motion, turbulence, and ambipolar drift. It also computes the charges carried by each grain species and the ion and electron densities, as well as the magnetic resistivities.  }
  % results. 
   {We find that the dust distribution significantly evolves during the protostellar collapse, shaping the magnetic resistivities. The peak size of the distribution,  population of small grains, and, consequently, the magnetic resistivities are controlled by both coagulation and fragmentation rates. Under standard assumptions, the small grains coagulate very early as they collide by ambipolar drift, yielding magnetic resistivities that are many orders of magnitude apart from the non-evolving dust case. In particular, the ambipolar resistivity, $\eta_\mathrm{AD}$, is very high prior to $n_\mathrm{H} = 10^{10} \ \mathrm{cm^{-3}}$. As a consequence, magnetic braking is expected to be ineffective. In this case, large size protoplanetary discs should result, which is inconsistent with recent observations. To alleviate this tension, we identified mechanisms that are capable of reducing the ambipolar resistivity during the ensuing protostellar collapse.  Among them, electrostatic repulsion and grain-grain erosion feature as the most promising approaches.}
  % conclusions. 
   {
   The evolution of the magnetic resistivities during the protostellar collapse and consequently the shape of the magnetic field in the early life of the protoplanetary disc strongly depends on the possibility to repopulate the  small grains or to prevent their early coagulation. Therefore, it is crucial to better constrain the collision outcomes and the dust grain's elastic properties, especially the grain's surface energy based on both theoretical and experimental approaches.
   }


    \authorrunning{Vallucci-Goy et al.}
    \titlerunning{Dust evolution during the protostellar collapse}

    \maketitle

%
%-------------------------------------------------------------------
\section{Introduction}

Interstellar dust plays a significant role in many astrophysical environments, in particular, in the context of star formation. Although it represents (on average) only $1\%$ of the diffuse  interstellar medium (ISM) mass, the wide spectrum of dust grain sizes affects the optical properties of the medium and, consequently, the heating and cooling processes at work \citep{McKee2007}. In addition, dust grains are at the heart of many molecule formation processes  \citep[including $\mathrm{H_2}$][]{Gould1963}. They are also the fundamental bricks of planets, which are believed to form rapidly in less than 1 Myr \citep{Testi2014}. The pathway to planetesimals and thereafter planets remains poorly understood. Getting a better description of the dust evolution during the protostellar collapse would allow us to provide more realistic dust initial conditions in protoplanetary discs. Finally, dust grains affect the ionisation degree of the medium and can become the main charge carriers during the protostellar collapse as they allow for electron and ion recombination and electron capture at their surface \citep{Marchand2016}. As such, they control the degree of coupling between the gas and the magnetic field \citep{Nakano2002} via the ambipolar, Hall, and ohmic magnetic resistivities, and, consequently, the evolution of the angular momentum through magnetic braking \citep{Mouschovias1991}. Indeed, dust plays a crucial role in shaping the magnetic field during the protostellar collapse and, thus, in influencing the structure of the resulting protoplanetary discs \citep{Wurster2016,Wurster2018,Hennebelle2020}.

The dust size spectrum is observed or at least modelled as a power-law Mathis-Rumple-Nordsiek (MRN) distribution in the diffuse ISM \citep{Mathis1977}. It was noticed that modifying the population of very small grains in the protostellar collapse simulations radically affects the magnetic resistivities and, consequently, the formation of protoplanetary discs \citep{Zhao2016,Marchand2016}. In addition, the dust size distribution is expected to evolve as dust grains undergo coagulation and fragmentation processes \citep{Dominik1997,Ormel2009}. Following such an evolution of the dust distribution is very challenging,  but it is essential to correctly describe the non-ideal MHD effects since the coupling between the grains, the gas, and the magnetic field depends directly on the total dust cross section. Indeed, \cite{Guillet2020}, \cite{Silsbee2020} and \cite{Lebreuilly2023} showed that a coagulating dust distribution would affect the resistivities during the protostellar collapse. So far, only a handful of studies have accounted for dust coagulation in 2D \citep{Vorobyov2019,Tu2022} or 3D \citep{Tsukamoto2021,Bate2022,Tsukamoto2023}. Recently, \cite{Marchand2023}  performed the first 3D simulations of the protostellar collapse and the early evolution of the protoplanetary disc, while including and  self-consistently computing the magnetic field from the magnetic resistivities. The dust coagulation (fragmentation was not included) was modelled at low computational cost with a turbulent kernel \citep[as described in][]{Marchand2021}, exerting an indirect feedback on the gas through the magnetic field, which allows us to better describe the effect of dust growth on non-ideal MHD effects. Similarly, \cite{Tsukamoto2023} used a dust single-size approximation and describe the co-evolution of dust grains and protoplanetary discs. Fragmentation has been introduced by \cite{Lebreuilly2023} and \cite{Kawasaki2022}, and some parameters (turbulent intensity and velocity threshold for fragmentation) was investigated in more detail in \cite{Kawasaki2023}. They show that the interplay between coagulation and fragmentation plays a critical role in regard to the evolution of the magnetic resistivities. Indeed, those studies showed that fragmentation could potentially repopulate the small grains at intermediate and high densities, consequently allowing the Ohm and Hall resistivities to rise. The ambipolar resistivity, $\eta_\mathrm{AD}$, was reduced to some extent prior to $n_\mathrm{H} < 10^{10} \mathrm{cm^{-3}}$ and enhanced thereafter. Besides, every study accounting for dust evolution produces magnetic resistivities, which remain remarkably away from the ones obtained with a non-evolving MRN distribution. This is even more striking when including ambipolar diffusion as a source of grain-grain collision, which is very efficient at depleting the smallest grains. For instance, as a consequence of small grain depletion by coagulation, the ambipolar resistivity, $\eta_\mathrm{AD}$, exhibits very high values at low and intermediate densities. This would have a profound impact on the properties of the resulting protoplanetary discs. Given that recent observational surveys, including CALYPSO \citep{Maury2019}, have shown that large discs ($ > 60 \ \mathrm{AU}$) are rare, serious doubts are cast on the likeliness of such high ambipolar resistivity values.

To shed light on the impact of dust evolution on the magnetic resistivities during the protostellar collapse, we conducted a thorough investigation of the influence of key parameters and physical effects through single-zone collapse numerical simulations. In particular, we investigated the role of dust grain properties (grain surface energy, intrinsic density, monomer size), collision outcomes (fragments distribution, maximum fragment size, grain-grain erosion, sticking efficiency, electrostatic repulsion), and the collapse properties (dust-to-gas ratio, turbulence intensity, ambipolar drift intensity, initial magnetic field strength, and ionisation rate induced by cosmic rays). However, we focus only on the most influential of the aforementioned parameters and strive to identify mechanisms that would allow us to mitigate the sharp rise in ambipolar resistivity, $\eta_\mathrm{AD}$.

In Sects. \ref{Sect dust model} and \ref{Section numerical methods}, we present the dust model and the numerical methods. Then, in Sect. \ref{Section results} and \ref{magnetic field section}, we share some of our results, which are discussed in Sect. \ref{Discussion}. Finally, Sect. \ref{Conclusion} provides concluding remarks and additional material can be found in the appendix.

\section{Dust model}
\label{Sect dust model}

This numerical work is carried out with the {\ttfamily {\ttfamily shark}} code. More details regarding the methods employed and the structure of the code can be found in the associated paper \citep{Lebreuilly2023}.

\subsection{Sources of dust relative velocities}
 \label{sec:sources}

Here, we present  the different sources of relative velocities that drive dust evolution. In the rest of the paper, the subscripts $i$ and $j$ will refer to the two colliding grains. 

\subsubsection{Brownian motion}

Dust grains are prone to Brownian motion, which is a thermal process that induces relative velocities as per: 

\begin{equation}
\Delta  v_{\mathrm{brownian},i,j} \equiv \sqrt{\frac{8 k_{\mathrm{B}} T}{\pi}} \sqrt{\frac{m_i + m_j}{m_i m_j} },
\end{equation}
where $T$ is the gas temperature and $k_{\mathrm{B}}$ the Boltzmann constant. This induced velocity is the highest for collisions including one grain much smaller than the other (or both small grains). Indeed in either case, the reduced mass in the square root is proportional to the mass of the smallest grain $\frac{1}{\mu} \propto \frac{1}{m_i}.$

\subsubsection{Ambipolar diffusion}

Ambipolar diffusion emerges as a consequence of different coupling between the magnetic field and dust grains, the said coupling depending on their size. This coupling is embodied by the gyromagnetic frequency $\omega_i = \frac{q_i B}{m_i}$. $q_i$ is the total charge carried by the grain, $m_\mathrm{i} = \frac{4}{3} \pi s_{i}^{3} \rho_\mathrm{grain}$ its mass, $s_\mathrm{i}$ its size and $\rho_\mathrm{grain}$ is the grain bulk intrinsic density, taken to be equal to $2.3 \ \mathrm{\gcc}$. $B$ is the magnetic field intensity. Following the approach proposed by \cite{Guillet2020}, the induced relative velocity between two grains is a function of their respective Hall factor, given by 

\begin{equation} \label{Hall}
    \Gamma_i = \omega_i t_\mathrm{s,i},
\end{equation}
where $t_\mathrm{s,i}$ is the stopping time, representing the coupling of a grain to the gas \citep{Epstein1924}: 

\begin{equation}
t_{\rm{s},i} \equiv \sqrt{\frac{\pi \gamma}{8}} \frac{\rho_{\rm{grain}} s_i}{\rho \mathrm{c_{\rm{s}}}},
\label{stopping}
\end{equation}
where $\rho$ and $c_{\mathrm{s}} \equiv \sqrt{\gamma P/\rho}$ are respectively the mass density of the gas and the sound speed. 
The number of charges carried by a grain increases slower with respect to grain size than the grain mass does. Therefore, larger grains are associated with lower gyromagnetic frequencies, namely, they are poorly coupled to the magnetic field. Besides, they are weakly coupled to the gas as well, as suggested by the size dependency of the stopping time. Nonetheless, the stopping time and the gyromagnetic timescale both rise with respect to the grain size, but the latter does so more rapidly. This leads in fine to lower Hall factors for larger grains.

As in \cite{Spitzer1941}, we define the ambipolar diffusion velocity as:
\begin{equation}
\label{AD vectorial form}
\vec V_{\mathrm{AD}} \simeq \frac{1}{|\vec{B}|^2}\frac{c^2 \eta_{\rm{AD}} ((\nabla \times\vec{B} )\times \vec{B})}{4 \pi},
\end{equation}
where $c$ is the speed of light, $\eta_{\mathrm{AD}}$ is the ambipolar resistivity (in unit $\mathrm{s}$). The curl of the magnetic field is approximated as $\frac{|\vec{B}|}{\lambda_\mathrm{J}}$, where $\lambda_\mathrm{J}$ is the Jeans length. Following \cite{Guillet2007,Guillet2020}, we  have: 
\begin{equation} \label{AD efficiency}
V_{\mathrm{AD}} \simeq  \delta_\mathrm{AD} \frac{c^2 \eta_{\mathrm{AD}}}{4 \pi \lambda_\mathrm{J}},
\end{equation}
where $\delta_\mathrm{AD}$ is the amibipolar drift intensity, taken equal to unity if not specified otherwise. We note that the larger the ambipolar resistivity, $\eta_\mathrm{AD}$, the stronger the ambipolar drift velocity. Finally, the ambipolar drift velocity between two dust grains is given by:  
\begin{equation}
\label{AD relative velocity}
\Delta  v_{\mathrm{ambipolar},i,j} \equiv  V_{\mathrm{AD}} \left|\frac{|\Gamma_i| }{\sqrt{1+\Gamma_{i}^2}}-\frac{|\Gamma_j| }{\sqrt{1+\Gamma_j^2}}\right|.
\end{equation}

Ambipolar drift allows for the largest grains to collect the smallest ones. Therefore, ambipolar diffusion is a process very effective at removing the smallest grains, but not at increasing significantly the maximum size of the  distribution \citep{Lebreuilly2023}. Importantly, we stress that the ambipolar resistivity, $\eta_\mathrm{AD}$, and the ambipolar drift intensity, $\delta_\mathrm{AD}$, are two different quantities, which should not be confused henceforth.

\subsubsection{Turbulence}
\label{turbulence}

The vast spectrum of vorticies and eddies induced in a turbulent cascade can generate differential velocities between dust grains \citep{voelk1980}. Indeed, a grain of given size couples with a variety of eddies and acquires a kick in the velocity space as a consequence. 
The analytical model used in this work is the one derived by \cite{Ormel2007}. We follow the approach of \cite{Guillet2020} and assume the injection timescale of the turbulence to be the free-fall timescale 
\begin{equation} \label{frefall timescale}
t_{\mathrm{L}} = t_{\mathrm{ff}} =\sqrt{\frac{3\pi}{32 \mathcal{G}\rho}},
\end{equation}
where $\rho$ is the gas density and $\mathcal{G}$ the universal gravitational constant. The dissipation timescale $t_{\eta}$ of the turbulence is then computed via the injection timescale and the Reynolds number \citep{Ormel2009}

\begin{equation}
t_{\eta}= t_{\mathrm{ff}}/\sqrt{\mathrm{Re}}\,,
\end{equation}
 where \begin{equation}
\mathrm{Re} = 6.2 \times 10^{7} \sqrt{\frac{n_{\mathrm{H}}}{10^{5} \ \mathrm{cm^{-3}}}} \sqrt{\frac{T }{10~\mathrm{K}}}.
\label{eq:Re}
\end{equation}

The Stokes number of a grain is defined as $\mathrm{St_i} = t_\mathrm{s,i}/t_\mathrm{ff}$, where the stopping time $t_\mathrm{s,i}$ of the largest of the two grains determines the three coupling regimes defined in the framework of \cite{Ormel2009}. The differential turbulent induced velocity writes: 

\begin{equation}
\Delta v_{\mathrm{turb,i,j}}^2 \equiv \left\{
    \begin{array}{ll}
       V_{\mathrm{g}}^2 \, \frac{\mathrm{St}_i-\mathrm{St}_j}{\mathrm{St}_i+\mathrm{St}_j} (\frac{\mathrm{St}_i^2}{\mathrm{St}_i+1/\sqrt{\mathrm{Re}}}+\frac{\mathrm{St}_j^2}{\mathrm{St}_j+1/\sqrt{\mathrm{Re}}}), \mbox{~if~} t_{\rm{s},i} < t_{\eta} \\
        V_{\mathrm{g}}^2  \, \beta_{\mathrm{i,j}} \mathrm{St}_i   \mbox{,~if~}  t_{\eta} \le t_{\rm{s},i} < t_{\mathrm{ff}} \\
      V_{\mathrm{g}}^2  \, (\frac{1}{\mathrm{St}_i+1}+\frac{1}{\mathrm{St}_j+1}) \mbox{, otherwise,}
        \end{array}
        \right.
\end{equation}
where $ V_{\mathrm{g}} = \sqrt{\theta}c_{\rm{s}}$  and $\beta_{\mathrm{i,j}}=3.2 -(1+x_{i,j})+\frac{2}{1+x_{i,j}}(\frac{1}{2.6}+\frac{x_{i,j}^3}{1.6+x_{i,j}})$, with $x_{j,i}$ being the ratio between $\mathrm{St}_j$ and $\mathrm{St}_i$. The parameter $ \theta $ is the turbulence intensity.  The usual value taken in the literature in the context of a gravitational collapse is $\theta=\frac{3}{2}$ \citep{Guillet2020}.

When the stopping time of the largest grain is comprised between the dissipation timescale and the injection timescale (i.e. turnover time of the largest eddies), the contribution from class II eddies acts as a random kick in the grain motion, allowing for  grains of similar size to efficiently collide \cite[see][for a detailed definition of the class I, II and III eddies]{Ormel2007}. This is the intermediate regime, which is the dominant regime in the context of the protostellar
collapse \citep{Marchand2021}.

We note that the turbulent source of relative velocities is commonly found to be the most efficient at growing large grains and significantly increasing the maximum size of the dust distribution \citep{Silsbee2020,Guillet2020,Lebreuilly2023}.

\subsection{Dust evolution}

The different sources of relative velocity presented here lead to grain collisions that may result in various outcomes. In this work, we only account for grain coagulation (perfect and imperfect sticking) and fragmentation, and we also investigate the influence of grain-grain erosion with a simplistic approach. Many other mechanisms have been identified with laboratory experiments, such as gas-grain erosion, abrasion, compaction and bouncing, mass transfer, and rotational disruption \cite[see][for more details]{Wurm2021}.  The grains are considered perfectly spherical, with a fixed and unique intrinsic density. 

\subsubsection{Coagulation}

The equation of mass conservation for the grains is expressed as: 
\begin{equation}
    \frac{\partial   \rho_{k} }{\partial t} +\nabla \cdot \left[ \rho_{k} \vec{v_k} \right] = S_{k,\mathrm{growth}}, \ \forall k \in \left[1,\mathcal{N}\right], 
\end{equation}
The subscript $k$ designates one specific grain among $\mathcal{N}$ species. Here, the term species relates to a grain of given size. The source term $S_{k,\mathrm{growth}}$ accounts for coagulation and fragmentation events and is computed according to the Smoluchowski equation \citep{Smolu16}:
\begin{equation}
S_{k,\mathrm{growth}} = \sum_{i+j\rightarrow k} K_{i,j} (m_i+m_j)\, n_{j} n_i -n_k m_k \sum_i^{\mathcal{N}} K_{k,i} n_i,
\end{equation}
where $K$ is the coagulation kernel that embodies the microphysics of the collision. The first term highlights the increase in mass density of the grain $k$ of mass $m_k \equiv m_i + m_j$ as a result of the sticking between two smaller grains $i$ and $j$. The second term (a sink term) describes the reduction in mass density of the grain $k$ for any sticking/coagulation it may undergo with an arbitrary grain $i$. $n_i$ is the number density of grains of mass $m_i$ such as $\rho_i \equiv m_i \,n_i$. 

The coagulation kernel is expressed as
\begin{equation}
K_{i,j} = \mathrm{max} \left(1 - \frac{E_\mathrm{Coulomb}}{E_\mathrm{Kinetic}},0 \right) \sqrt{\frac{8}{3\pi}}   \pi (s_{i} + s_{j})^2 \Delta  v_{i,j},
\label{kernel}
\end{equation}
 where $ \Delta  v_{i,j}$ is the differential velocity between grains $i$ and $j$, of size $s_i$ and $s_j$, defined as the quadratic sum of the sources of differential velocity detailed in Sect. \ref{sec:sources}. The $\sqrt{\frac{8}{3\pi}}$ pre-factor comes from considering that grains relative velocities along the three x, y, and z axis are Gaussian variables of variance $\Delta v_{i,j}^2/3$ \citep{Guillet2020,Marchand2021}. Here, $E_\mathrm{kinetic} = \frac{1}{2} \mu_{i,j}  \Delta  v_{i,j}^2 $ is the kinetic energy of the colliding grains. $\mu_{i,j} \equiv m_i m_j/\left(m_i+m_j\right) $ is the reduced mass of the two grains. Also, $E_\mathrm{Coulomb} = q_\mathrm{i} q_\mathrm{j}/4 \pi \epsilon_\mathrm{vacuum} \left(s_\mathrm{i} + s_\mathrm{j} \right)$ is the Coulomb electrostatic energy and $\epsilon_\mathrm{vacuum}$ the vacuum dielectric permittivity. The first term in the Kernel accounts for the cross section modification induced by the electrostatic interaction between the two colliding grains \citep{Akimkin2023}. It is included only in the simulations for which electrostatic repulsion is explicitly investigated.

\subsubsection{Fragmentation and elastic properties}
\label{sec:frag}

To establish a fragmentation condition, we follow \cite{Ormel2009}, whose work involves dust grains in quiescent environments, that is to say: porous aggregates colliding at subsonic velocities and made up of elementary, unbreakable pieces called monomers. Experimentally \cite{Blum2000} observed that complete fragmentation of the two colliding grains takes place as soon as the kinetic energy of the collision exceeds five times the total breaking energy, $E_\mathrm{br,tot} = N_\mathrm{tot}E_\mathrm{br}$, of the grains \cite[defined in][]{Dominik1997}:

\begin{equation} \label{fragmentation condition}
\frac{1}{2} \mu_{i,j}  \Delta  v_{i,j}^2 > 5 N_{\mathrm{tot}}  E_{\mathrm{br}},
\end{equation}

\begin{equation}
E_{\mathrm{br}} = A_{\mathrm{br}} \gamma_{\mathrm{grain}}^{5/3}\frac{(s_{\mathrm{mono}}/2)^{4/3}}{\varepsilon_{*}^{2/3}},
\label{braking nrj}
\end{equation}
where $N_{tot} \equiv \frac{m_i+m_j}{m_\mathrm{mono}}$ is the total number of electrostatic bonds between monomers. $E_\mathrm{br}$ is the energy required to break a single electrostatic bound and depends on the elastic properties of the grains, including the surface energy density $\gamma_\mathrm{grain}$ and the reduced elastic modulus $\varepsilon_{*}$. The latter is a function of the Poisson's ratio and Young's modulus, as detailed in \cite{Dominik1997}. The mass of a monomer is related to the average monomer size as per $m_\mathrm{mono}\equiv \frac{4}{3} \pi s^{3}_\mathrm{mono} \rho_\mathrm{grain}$ where $\rho_\mathrm{grain}$ is the bulk intrinsic density of the grains. We use $A_{\mathrm{br}}=2.8 \times 10^{3}$, which is experimentally measured in \cite{Blum2000}. Here,  $s_\mathrm{mono}$ is the monomer size, taken to be equal to $100 \ \mathrm{nm}$. We note that although $E_\mathrm{br}$ diminishes when reducing the monomer size, the total breaking energy actually increases due to the increasing number of bonds $N_\mathrm{tot}$. Consequently, reducing the monomer size leads to a higher fragmentation energy threshold.

\section{Numerical methods}
\label{Section numerical methods}

The numerical structure of the code used to treat the dust evolution has been thoroughly detailed in \cite{Lebreuilly2023}. Here, we only shed light on the specific features relevant to our analysis. 
The version used here is the python {\ttfamily sharkpy} version.
It accounts for dust evolution through coagulation and fragmentation processes, and computes the grain charge, ion, and electron number density as well as the magnetic conductivities and resistivities of the mixture.
Within the scope of this work, the simulations were performed on a single cell, where physical properties can evolve with time, but with no spatial extension.

\subsection{Dust}

\subsubsection{Coagulation and fragmentation}
\label{coag and frag}

Unless otherwise specified, the total dust-to-gas ratio is taken to be  $\epsilon_0=0.01$, and is spread over $\mathcal{N}$ grain sizes (i.e. bins or species) on a logarithmic grid that ranges between $s_{\mathrm{min}}$ and $s_{\mathrm{max}}$. The number of dust bins $\mathcal{N}$ used in regard to the dust growth algorithm is inferred from the logarithmic increment, and the limits of the grid. The logarithmic increment writes: $\zeta_\mathrm{log}= \left ( \frac{s_\mathrm{max}}{s_\mathrm{min}}\right )^{\left( \frac{1}{\mathcal{N}}\right)}$. If the grid limits are to be modified, the logarithmic increment (i.e. the number of dust bins per decade) is held constant and we adjust the total number of dust bins accordingly, with the reference value $\left( s_\mathrm{min} = 5 \times 10^{-7} \ \mathrm{cm} , s_\mathrm{max} = 1 \ \mathrm{cm}\right) \Rightarrow \mathcal{N}=300$ extracted from a convergence test.
 The initial dust distribution is considered to be a power-law as: $\frac{dN}{ds} = s^{\lambda}$ for the number density , or equivalently: $\frac{d \epsilon}{d \mathrm{log(s)}} = s^{\lambda + 4}$ in terms of mass dust-to-gas ratio, with $    \int_{s_\mathrm{min}}^{s_\mathrm{max}}\frac{d\epsilon}{dlog(s)}dlog(s) = \epsilon_0$ and $\lambda=-3.5$.

For each collision between grains i and j, we define a fragmentation coefficient $f_{\mathrm{frag},i,j}$ that is computed according to the fragmentation condition presented in Sect. \ref{sec:frag} (and explained below). A fraction
$(1-f_{\mathrm{frag},i,j})\, ( m_i +m_j)$ of the total mass involved in the collision is coagulated to form a larger grain. The rest of the mass, $f_{\mathrm{frag},i,j}\, ( m_i +m_j) = m_\mathrm{frag}$, is shattered and redistributed as fragments among all the bins in the range $\left[m_\mathrm{min},0.1 \ m_\mathrm{frag} \right]$, with $m_\mathrm{min}$ being the mass associated with the dust bin of size $s_\mathrm{min}$.  The shattered mass $m_\mathrm{frag}$ is redistributed as a number density power-law: $\frac{dN}{ds} = s^{\psi}$, where $\psi = -3.5$.

We set $f_\mathrm{frag} = 1$ when the fragmentation condition in Sect. \ref{sec:frag} is met (corresponding to a complete fragmentation) and  $f_\mathrm{frag} = 0$ when $\frac{1}{2} \mu_{i,j}  \Delta  v_{i,j}^2 < 0.1 E_\mathrm{br,tot}$ (corresponding to sheer coagulation). In between those limits, the fragmentation coefficient varies linearly with respect to the kinetic energy of the collision. 

\subsubsection{Grain-grain erosion}

For some simulations, we constrain the fragmentation coefficient to be finite ($0 < f_\mathrm{frag,min} = f_\mathrm{ero}$) when the fragmentation criterion is not met, in such a way as to include a soft grain-grain erosion \citep{Blum2018}.

Grain-grain erosion involves grains of very different sizes. Recently, \cite{Schrapler2018} conducted laboratory experiments to constrain the velocity threshold (for the onset of grain-grain erosion) and the corresponding efficiency by colliding micrometer projectiles onto a cm size target. They measured mass losses from the larger grain equal to a few times the mass of the smaller one in some cases. They observed the velocity threshold decreasing and the erosion efficiency rising when increasing the size difference between the projectile and the target. We assume their results can be translated to different grain size regimes (e.g. nanometer and micrometer grains at low density during the collapse). We implemented in our code the erosion velocity threshold (Fig. 8 in their paper) as a function of the grain size ratio $s_\mathrm{ratio}= s_\mathrm{target}/s_\mathrm{projectile}$ (where $s_\mathrm{target}$ is the size of the larger grain and $s_\mathrm{projectile}$ that of the smaller) 
\begin{equation}
    \mathrm{log} \left(V_\mathrm{erosion} \right) =  \frac{\mathrm{log}\left(3 \right)}{\mathrm{log}\left(0.2\right)} \mathrm{log}\left(s_\mathrm{ratio}\right) + \mathrm{log} \left(45 \right) - 2 \frac{\mathrm{log}\left(3\right)}{\mathrm{log}\left(0.2\right)} \ \left[\mathrm{m \ s^{-1}}\right],
\end{equation}
and the erosion efficiency (Fig. 7 in their paper):

\begin{equation}
\label{erosion efficiency equation}
    f_\mathrm{ero} =  \left(\frac{1}{s_\mathrm{ratio}} \frac{s_\mathrm{target}}{2 \times 10^{-5} \ \mathrm{m}} \right)^{-0.62} \frac{\Delta v}{15 \ \mathrm{m \ s^{-1}}} - 1,
\end{equation}
where $\Delta v$ is the relative collision velocity between both grains.

\subsubsection{Electrostatic repulsion}

Since grains are negatively charged on average, they experience electrostatic repulsion as they encounter, which modifies their effective collisional cross-section (see Eq. \ref{kernel}). In addition, we allow for coagulation only if the kinetic energy of the colliding grains is higher than the Coulomb electrostatic energy ($E_\mathrm{kinetic} > E_\mathrm{Coulomb}$). This translates into a velocity threshold that needs to be exceeded for the colliding grains to stick together:  
 \begin{equation} \label{Coulomb barrier}
     \Delta v_{i,j} > \Delta v_\mathrm{Coulomb,threshold} = \sqrt{\frac{q_i q_j}{2 \pi \epsilon_{\mathrm{vacuum}} \left(s_i + s_j \right) \mu_\mathrm{i,j}}}.
 \end{equation} 
This velocity condition is included only in the simulations for which electrostatic repulsion is explicitly investigated.

\subsubsection{Charging and resistivities}
\label{charging and res section}

We used the scheme presented in \cite{Marchand2021} to compute the charge of each grain species and the number density of ions and electrons. It relies on different  parameters such as the average ion ($\mathrm{HCO^{+}}$) mass $\mu_\mathrm{ions} = 25$ , the size of ice coats $s_\mathrm{ice} = 8 \ \mathrm{nm}$, the sticking efficiency coefficient of electrons onto grains $s_\mathrm{e} = 0.5$ and the cosmic-ray ionisation rate  $\zeta = 5 \times 10^{-17} \mathrm{s^{-1}} $.  

 The conductivities (parallel, perpendicular and Hall) are inferred via the Hall factor (Eq. \ref{Hall}) and the individual conductivities of the different species: 

\begin{align} \label{conductivities expressions}
\sigma_{\mathrm{par}}&= \sum_j \sigma_j, \nonumber \\ 
\sigma_{\mathrm{perp}}&=  \sum_j  \sigma_j \frac{1}{1+\Gamma_j^2} , \nonumber \\ 
\sigma_{\mathrm{H}}&=- \sum_j  \sigma_j \frac{\Gamma_j}{1+\Gamma_j^2}.
\end{align}
In turn, the ohmic, ambipolar and Hall resistivities are computed as 

\begin{align} \label{resistivities expressions}
\eta_{\mathrm{O}}&= \frac{1}{\sigma_{\mathrm{par}}}, \nonumber \\ 
\eta_{\mathrm{AD}}&=\frac{\sigma_{\mathrm{perp}}}{\sigma_{\mathrm{perp}}^2+\sigma_{\mathrm{H}}^2} -\frac{1}{\sigma_{\mathrm{par}}} , \nonumber \\ 
\eta_{\mathrm{H}}&= \frac{\sigma_{\mathrm{H}}}{\sigma_{\mathrm{perp}}^2+\sigma_{\mathrm{H}}^2}.
\end{align}
To better understand the behavior of the resistivities,  in  Appendix \ref{What species control the conductivities ?}, we include the conductivity profiles, with an emphasis on the distinct influence of the different species, namely, the electrons, the ions and the grains. 

\subsection{Setups}

\subsubsection{Magnetic field}

We use an analytical expression that allows for the magnetic field to increase with the density as per \cite{Li2011} and \cite{Marchand2016}: 

\begin{equation} \label{mag field equation}
 B = B_0 \sqrt{ \frac{n_{\mathrm{H}}}{10^{4} \ \mathrm{cm^{-3}}}},
\end{equation}
where $B_0 = 30 \ \mu \mathrm{G}$ is the magnetic field intensity at $n_\mathrm{H} = 10^4 \ \mathrm{cm^{-3}}$. We define the ambipolar diffusion timescale as:
\begin{equation} \label{AD timescale}
t_{\mathrm{AD}} =\frac{4\pi}{c^2}\frac{r^2}{\eta_{AD}},
\end{equation}
where $c$ is the speed of light and $r$ the distance from the center of the collapsing cloud.

\subsubsection{Collapse setup}

\begin{figure}
\centering
\includegraphics[width=0.5\textwidth]{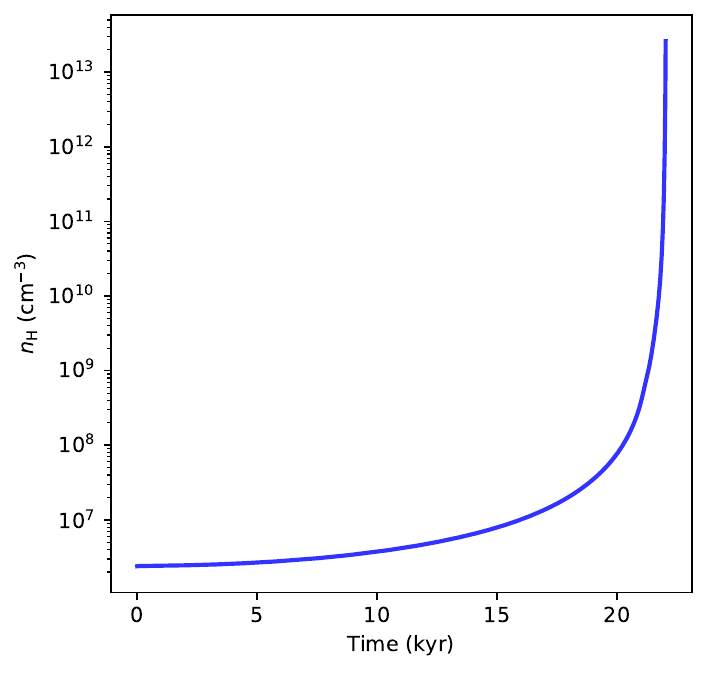}
\caption{Gas numerical density evolution with time throughout the gravitational collapse.}
\label{fig:nH_evolution}

\end{figure}

This single zone setup reproduces the gravitational collapse of a protostellar cloud up to the formation of the first Larson core. 

The temperature is computed according to a barotropic equation used in \cite{Machida2006}, \cite{Marchand2016} and \cite{Marchand2021}: 
\begin{equation}
    T = T_0 \sqrt{1+\left(\frac{n_\mathrm{H}}{n_\mathrm{H,ad}}\right)^{0.8}}\left(1+\left(\frac{n_\mathrm{H}}{10^{16} \ \mathrm{cm^{-3}}}\right)\right)^{-0.3},
    \label{barotropic}
\end{equation}
where the initial cloud temperature $T_0 = 10 \ \mathrm{K}$ is a typical value measured in dense prestellar cores. Also, $n_\mathrm{H,ad}$ is the threshold density that marks the transition between the isothermal and the adiabatic regime,  usually $n_\mathrm{H,ad} \simeq 10^{11} \ \mathrm{cm^{-3}}$. 
We consider a perfect gas with equation of state: 
\begin{equation}
P_\mathrm{g} = \frac{\rho k_B T}{\mu_\mathrm{g} m_\mathrm{H}},    
\end{equation}
where $P_\mathrm{g}$ is the gas pressure, $\rho$ is the mass density of the gas, $\mu_\mathrm{g} = 2.3$ is the mean molecular weight,  and $m_\mathrm{H}$ is the mass of a hydrogen atom. The sound speed is given by $c_{\mathrm{s},0} \equiv \sqrt{\frac{\gamma P_\mathrm{g}}{\rho}} = \sqrt{\frac{\gamma k_{\mathrm{B}}T_0}{\mu_{\rm{g}} m_{\rm{H}}}}$, where the effective adiabatic index is taken to be that of a monoatomic gas  $\gamma = \frac{5}{3}$ since the temperature of $H_2$ molecules is not high enough to excite vibrational and rotational energy states in this cold environment.
Figure \ref{fig:nH_evolution} gives the gas density evolution with time during the protostellar collapse. The data is extracted from the central cell of a {\ttfamily shark} 1D collapse simulation without rotation or magnetic support.
The collapse is initiated and halted exactly as in \cite{Lebreuilly2023}, that is, to say with an initial density of $\rho = 9.2 \times 10^{-18} \ \mathrm{g \ cm^{-3}} \left ( n_\mathrm{H} = 2.4 \times 10^{6} \ \mathrm{cm^{-3}} \right)$ and a final density of $\rho = 10^{-10} \ \mathrm{g \, cm^{-3}} \left (n_\mathrm{H} = 2.6 \times 10^{13} \ \mathrm{cm^{-3}}\right)$, which corresponds to a solar mass cloud associated with a  gravitational to thermal energy ratio $\alpha = 0.25$.

In Appendix \ref{coag_frag_comparison}, we offer a comparison between a non-evolving MRN dust distribution, a growing distribution and a growing plus fragmenting distribution at different instants of the collapse. We recovered the results of \cite{Lebreuilly2023} and \cite{Kawasaki2022}.

\subsubsection{Parameters investigated}

Table \ref{table1} displays the parameters explored and their corresponding values. We highlight in bold in the first column the reference values used when a given parameter is not being varied (note: we used two reference values for the grain surface energy $\gamma_\mathrm{grain}$). From those reference values, we vary the parameters independently, one at a time.

    1. $\gamma_\mathrm{grain}$  is the surface energy of the grains, which controls the breaking energy. Given the uncertainties around the elastic properties and tensile strength of the grains, and given that there is a temperature dependency of the surface energy of the grains, we vary the latter between two extreme values measured in \cite{Musiolik2019}: $\gamma_\mathrm{grain} = 2.9 \ \mathrm{erg \ cm^{-2}}$ and  $\gamma_\mathrm{grain} = 190 \ \mathrm{erg \ cm^{-2}}$. Using the elastic modulus for icy-grains to be $\varepsilon_{*} = 3.7 \times 10^{10}$, the two extreme values for the surface energy correspond respectively to the case of minimum breaking energy ($E_\mathrm{{br,min}} = 1.27 \times 10^{-10} \ \mathrm{erg}$) and maximum breaking energy ($E_\mathrm{{br,max}} = 4.11 \times 10^{-7} \  \mathrm{erg}$) we could encounter.

    2. Erosion is a boolean that activates or deactivates the grain-grain erosion process.

    3. $\mathrm{Repulsion}$ is a boolean that activates or deactivates the grain-grain electrostatic repulsion process.

    4. $\delta_\mathrm{AD}$ is the ambipolar drift intensity, taking values between 0.1 and 10. Those limit values are expected in collapsing environments, as suggested by Fig. \ref{delta in PPD}.
    
    5. $B_{0}$ is the initial magnetic field strength corresponding to a gas density of $n_\mathrm{H} = 10^4 \ \mathrm{cm^{-3}}$ in the scaling approximation of Eq. (\ref{mag field equation}). It takes values between $10^{-5} \ \mathrm{G}$ and $5 \times 10^{-5} \ \mathrm{G}$, which are common values measured in dense cores, see for example \cite{Troland2008}.

    6. $\zeta$ is the ionisation rate induced by cosmic rays. (see Sect. \ref{charging and res section}). It takes values between $5 \times 10^{-18} \ \mathrm{s^{-1}}$ and $5 \times 10^{-16} \ \mathrm{s^{-1}}$, which are typical values measured in \cite{Padovani2022}. Constraining this parameter is a daunting task due to the uncertainties in the chemical networks used and to observation difficulties.

    7. $\epsilon_\mathrm{stick}$ is a grain sticking coefficient which affects the sticking probability between two colliding grains, but not that of fragmentation. It takes values between $0.1$ and $1$. Since this parameter is poorly constrained and dependent on the grain surface energy and grain coats nature, we explored a wide range of values.

\section{Results}
\label{Section results}

\begin{figure*}
\centering
    \includegraphics[width= 1\textwidth]{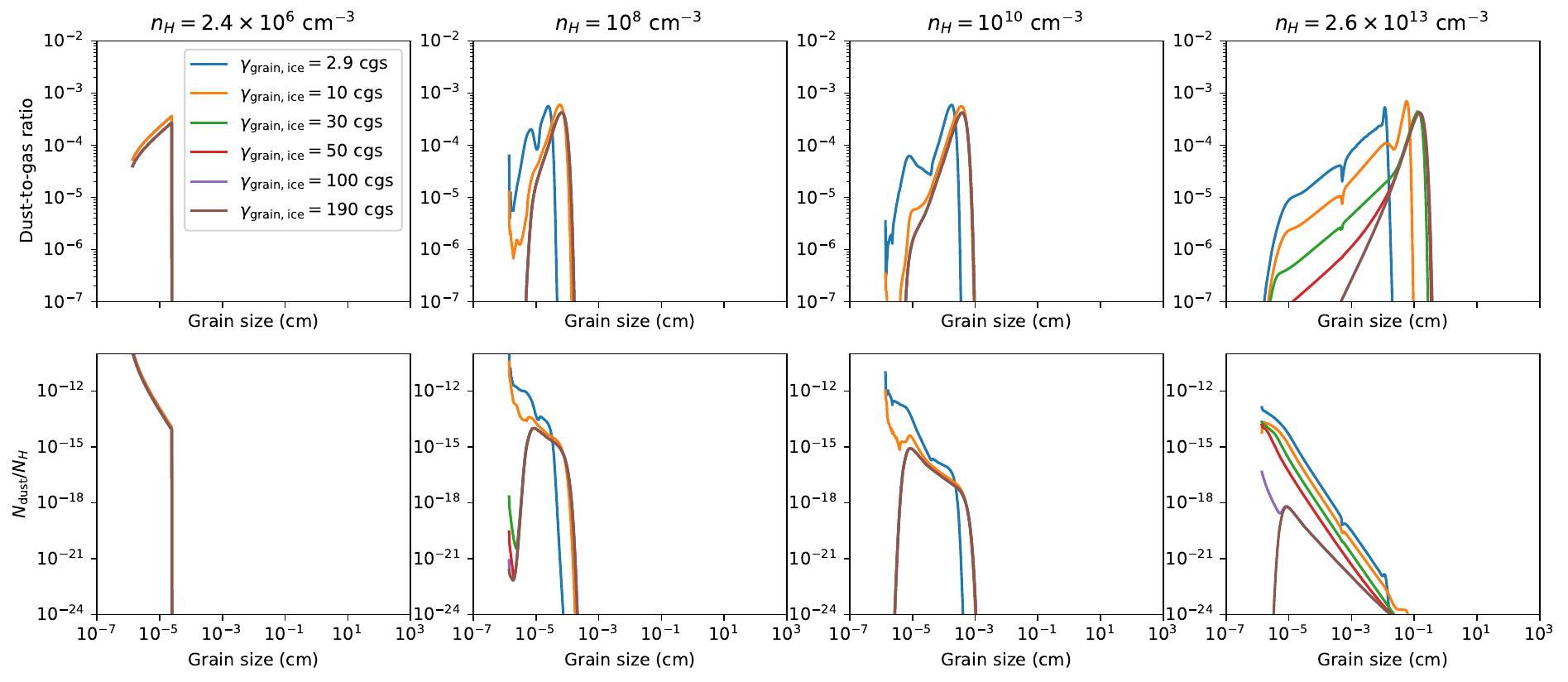}
    \includegraphics[width= 1\textwidth]{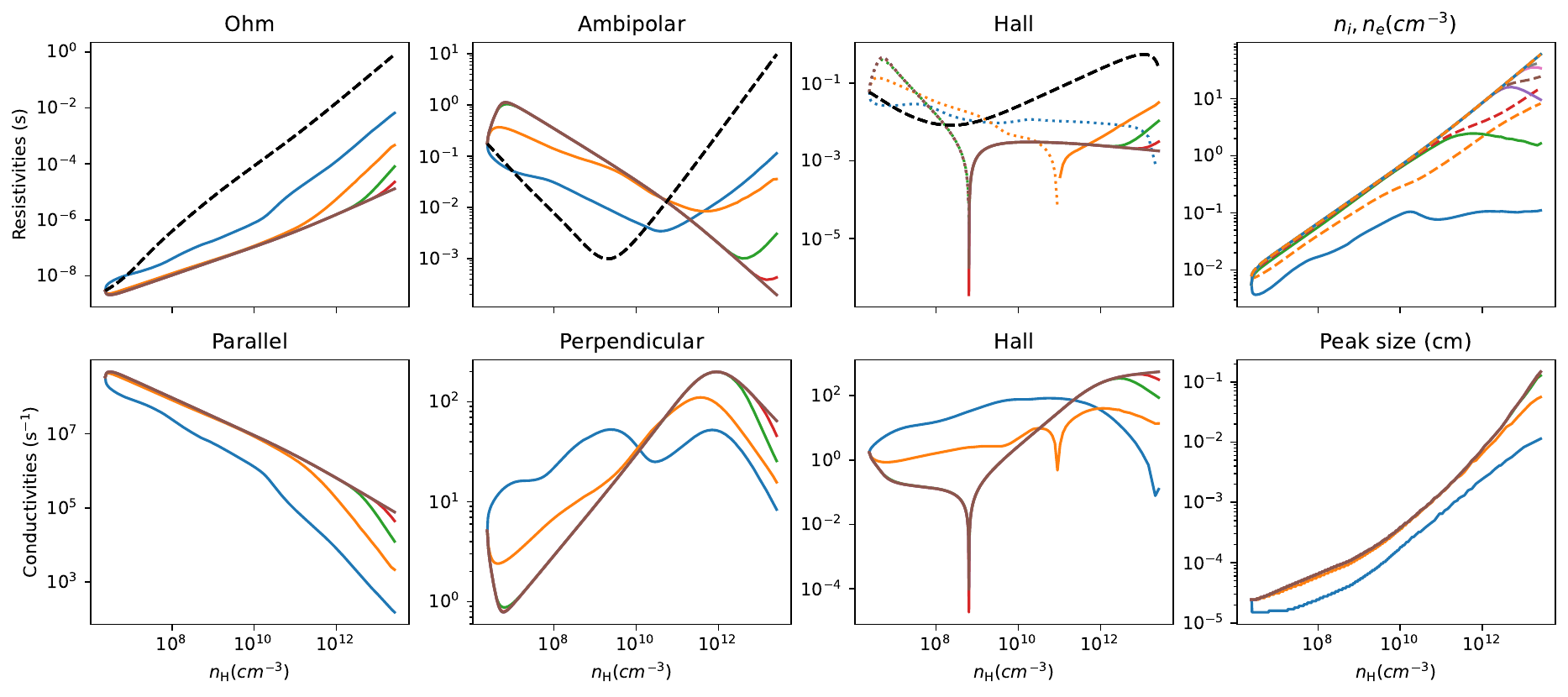}
    \caption{Dust distributions at different stages during the collapse. First row: Mass dust-to-gas ratio. Second row: Number dust-to-gas ratio. Third row: Resistivities profile and ion and electron numerical densities (Hall resistivity: dotted line for the negative values and solid line for the positive ones). Dashed black lines: Reference resistivities obtained with a fixed MRN dust distribution. Fourth row: Parallel, perpendicular, and Hall conductivities, and the evolution of the peak size of the dust distribution. Comparison between different values of the grain surface energy expressed in cgs units ($\mathrm{erg \ cm^{-2}}$).} 
    \label{res_gamma_comparison_icenopermfrag}
\end{figure*}

\begin{figure*}[hbt]
    \includegraphics[width= 1.0\textwidth]{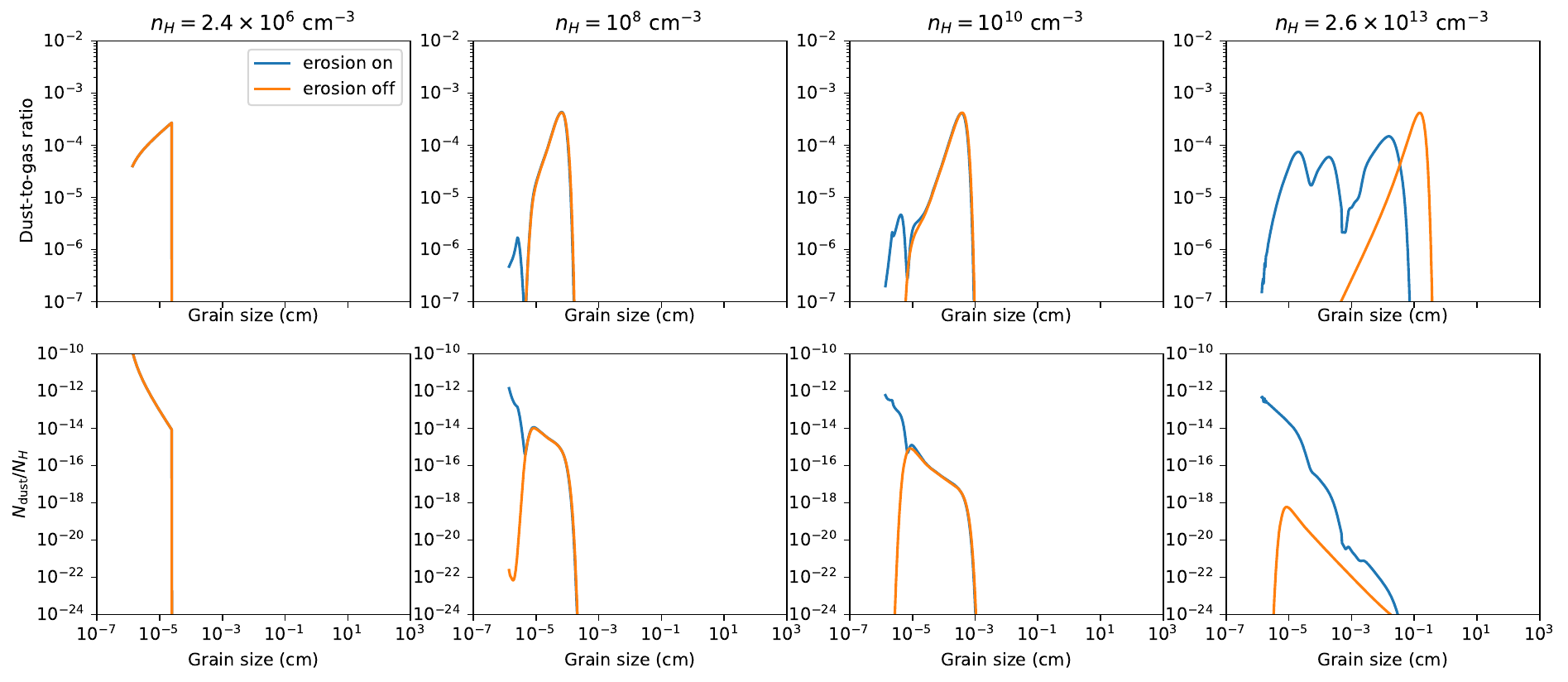}

    \includegraphics[width= 1.0\textwidth]{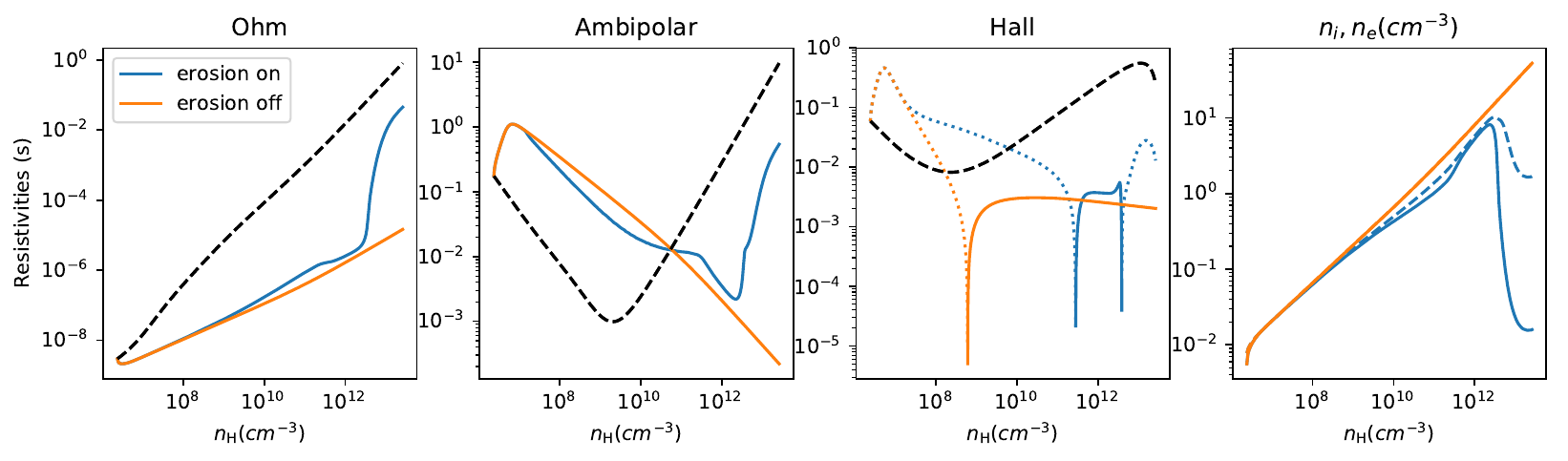}

    \caption{Same as Fig. \ref{res_gamma_comparison_icenopermfrag}, but investigating the influence of grain-grain erosion with $\gamma_\mathrm{grain} = 190 \ \mathrm{erg \ cm^{-2}}$.} 
    \label{res_erosion}
\end{figure*}

\begin{figure*}[hbt]
    \sidecaption
    \includegraphics[width=12cm]{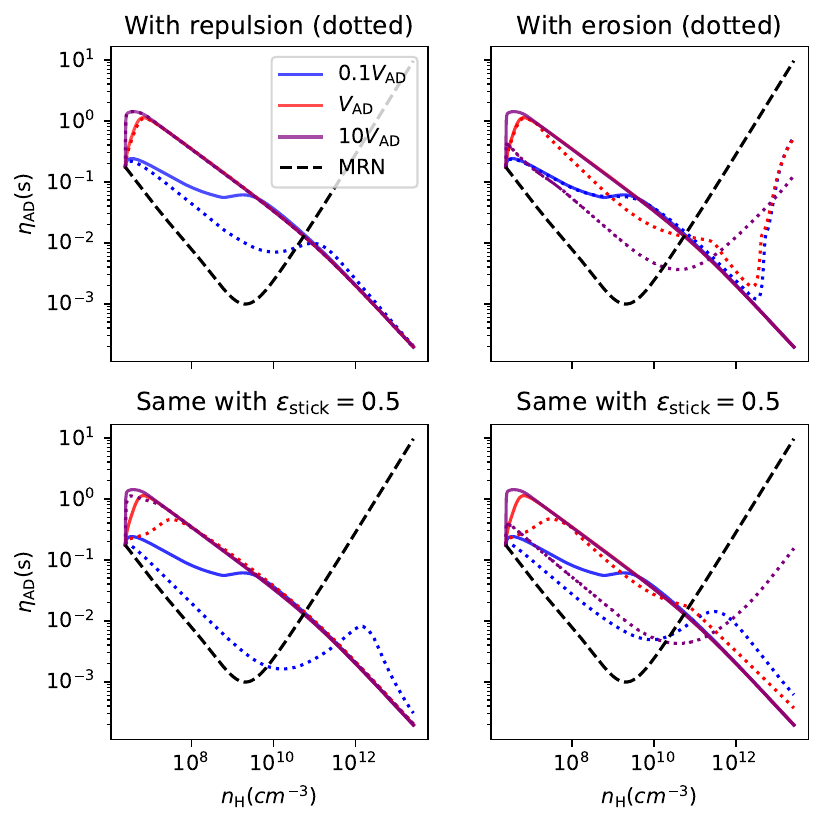}
    \caption{Ambipolar resistivity $\eta_\mathrm{AD}$ evolution during the protostellar collapse for different values of the ambipolar drift intensity or equivalently different ambipolar drift velocities: $\delta_\mathrm{AD} = 0.1 \  (0.1V_\mathrm{AD})$, $\delta_\mathrm{AD} = 1 \ (V_\mathrm{AD})$, and $\delta_\mathrm{AD} = 10 \ (10V_\mathrm{AD})$.  Grain-grain electrostatic repulsion (first column) and erosion (second column) are also included (dotted lines). Second row: Same details, but with a sticking efficiency of 0.5.}
    \label{res_electrostatic_repulsion+delta}
\end{figure*}

\begin{table}
\caption{Range of values taken by the explored parameters.}
\begin{center}

\begin{tabular}{ |p{3cm}||p{4cm}|}
 \hline
 \multicolumn{2}{|c|}{Parameter exploration} \\
 \hline
 $\gamma_\mathrm{grain} (\mathrm{erg \ cm^{-2}}) $&$[2.9,\boldsymbol{10},30,50,100,\boldsymbol{190}]$\\

  $\mathrm{erosion}$&$[\boldsymbol{\mathrm{False}},\mathrm{True}]$\\
  $\mathrm{repulsion}$&$[\boldsymbol{\mathrm{False}},\mathrm{True}]$\\

  $\delta_\mathrm{AD}$&$[0.1,\boldsymbol{1},10]$\\

 \hline
\hline

$B_0 (\mathrm{\mu G})$&$[10,\boldsymbol{30},50]$\\
$\zeta (\mathrm{s^{-1}})$&$[5 \times 10^{-18},\boldsymbol{5 \times 10^{-17}},5 \times 10^{-16}]$\\
 \hline
 \hline

$\epsilon_\mathrm{stick}$&$[0.1,0.5,0.7,\boldsymbol{1}]$\\
\hline

\end{tabular}

\tablefoot{In bold are the reference values used when a given parameter is not being varied. Two reference values have been used for the grain surface energy, corresponding to fragile grains on one hand, and to resilient grains on the other.}
\end{center}

\label{table1}
\end{table}

  We first explored the impact of the surface energy of the grains. Then, we found that the degree of influence of the other parameters depends strongly on the fragmentation rate, that is on the grain surface energy. Consequently, we do not make any assumption regarding the composition of the grains but rather choose to investigate the influence of the other parameters for two different extreme values of the grain surface energy, corresponding to resistant grains ($\gamma_\mathrm{grain} = 190 \ \mathrm{erg \ cm^{-2}}$) on one hand, and fragile ones ($\gamma_\mathrm{grain} = 10 \ \mathrm{erg \ cm^{-2}}$) on the other hand. We investigate the influence of the ambipolar drift intensity and its combined effect  with electrostatic repulsion and grain-grain erosion for a grain surface energy of $\gamma_\mathrm{grain} = 190 \ \mathrm{erg \ cm^{-2}}$. The analysis of the grain-grain sticking efficiency $\epsilon_\mathrm{stick}$ is given in the
  appendix (Appendix \ref{res_sticking_efficiency}),  performed with $\gamma_\mathrm{grain} = 190 \ \mathrm{erg \ cm^{-2}}$. The initial magnetic field strength $B_0$ and the cosmic-ray ionisation rate $\zeta$ are mentioned and commented on in Sect. \ref{magnetic field section} ($\gamma_\mathrm{grain} = 190 \ \mathrm{erg \ cm^{-2}}$ if not specified otherwise).

\subsection{Impact of the surface energy of the grains $\gamma_\mathrm{grain}$}

The surface energy is the most influential grain property in regard to fragmentation, since the breaking energy scales as $E_\mathrm{br} \propto \gamma_\mathrm{grain}^{\frac{5}{3}}$ (see Eq. \ref{braking nrj}). As a reminder, the correct value is still subject to debate in the literature.

Figure \ref{res_gamma_comparison_icenopermfrag}  depicts the evolution of the dust mass distribution (first row) and dust number distribution (second row) at different densities during the protostellar collapse, for different values of the grain surface energy. The third row displays the Ohm, ambipolar, Hall resistivities, ion and electron numerical densities evolution with gas density. Likewise for the last row, but with the conductivities and the peak size of the dust distribution. The magnetic resistivities associated with a fixed MRN dust distribution are displayed as a reference in black dashed lines. The impact of the grain surface energy is striking. Indeed, for values of $\gamma_\mathrm{grain} = 10 \ \mathrm{erg \ cm^{-2}}$ and below, fragmentation comes in very early as a replenishment of the population of small grains. It is observed already at $n_\mathrm{H} = 10^{8} \ \mathrm{cm^{-3}}$. Conversely for larger values, fragmentation only appears at densities higher than $n_\mathrm{H} = 10^{10} \ \mathrm{cm^{-3}}$, and is even completely absent for $\gamma_\mathrm{grain} = 190 \ \mathrm{erg \ cm^{-2}}$. The peak size of the distribution is lower when decreasing the surface energy: indeed at the end of the collapse, the peak size is of the order of $0.1 \ \mathrm{mm}$ for the lowest value of $\gamma_\mathrm{grain}$ while it can reach a value of $1 \ \mathrm{mm}$ in the case of the largest grain surface energies. In summary, a lower grain surface energy thus entails more fragile grains, which leads to a more rapid and efficient replenishment of the population of small grains along with a reduced peak size of the dust distribution at the end of the collapse.

Such variations in the population of small grains at low densities lead to significant variations in the resistivities. Indeed, for $\gamma_\mathrm{grain} = 2.9 \ \mathrm{erg \ cm^{-2}}$, the immediate replenishment of small grains at all densities gives rise to a subsequent much higher Ohm resistivity throughout the collapse, even at low densities. For $\gamma_\mathrm{grain} = 10 \ \mathrm{erg \ cm^{-2}}$, fragmentation occurs early but slightly later than $\gamma_\mathrm{grain} = 2.9 \ \mathrm{erg \ cm^{-2}}$, and consequently the  effects on the Ohm resistivity only emerge later on, as we need the density to sufficiently increase in order for the excess of small grains to collect the electrons. Indeed, a higher number of small grains implies a larger overall cross-section for electron capture by dust grains. This capture reduces the electron density and thus enhances the ohmic resistivity. The ambipolar resistivity $\eta_\mathrm{AD}$ is sensitive to the population of  small grains as well. In reducing the value of the surface energy of the grains, the small grain re-population results in a higher dust conductivity contribution (which, along with the ions, is the prevailing contribution here; see Appendix \ref{What species control the conductivities ?}) and thus to a lower ambipolar resistivity $\eta_\mathrm{AD}$ at low and intermediate densities. At high densities, the ambipolar resistivity is controlled by the ion density only, and the ambipolar resistivity appears as larger since the excess of small grains reduces the ion density by means of ion-electron recombination at their surface. Finally, the Hall resistivity is very sensitive to the population of small grains and reducing the surface energy induces a re-population of the latter, leading to a transition between negative and positive values that is shifted to higher densities.  This transition is even completely suppressed for the lowest value of $\gamma_\mathrm{grain}$, which displays a Hall resistivity profile quite similar to that of a fixed MRN distribution (represented as a black dashed line). 
Indeed, the Hall resistivity is very sensitive to the relative number of electrons and ions. A change in the number of small grains induces a change in the relative number densities of electrons and ions which can easily lead to a change of sign of the Hall resistivity. The more numerous the small grains, the later the transition (change of sign) occurs (see Sect. \ref{What species control the conductivities ?} for a better description of the Hall conductivity sensitivity to the dust distribution). Depending on the sign of the Hall resistivity, the Hall effect can either promote or reduce the rotation of the collapsing core.

Since the ambipolar resistivity $\eta_\mathrm{AD}$ dominates over the Ohm resistivity, we can conclude that more fragile grains (i.e. enhanced fragmentation) leads to a less efficient magnetic diffusion at low and intermediate densities ($n_\mathrm{H} \lesssim 10^{10} \ \mathrm{cm^{-3}}$), but more efficient at high density.

\subsection{Exploration for $\gamma_\mathrm{grain} = 190 \ \mathrm{erg \ cm^{-2}}$}

We go on to consider the case of resilient grains for which fragmentation is substantially curtailed.

\subsubsection{Impact of grain-grain erosion }

\label{erosion result section}

Here, we explore the impact of grain-grain erosion that can serve as an alternative to fragmentation to repopulate the small grains. We note on Fig. \ref{res_erosion} the replenishment of the population of very small grains induced by erosion and the limited impact it has on the magnetic resistivities at low density. However, the said impact gets stronger with density as the largest grains grow, allowing for larger grain size ratios and collisional velocities. We recall that according to Eq. (\ref{erosion efficiency equation}), the erosion efficiency augments for increasing grain size ratio and grain collisional velocity. In effect, while the model without erosion exhibits a clearly diminishing dust number distribution as the collapse proceeds, the model with erosion maintains a constant population of small grains, with a value hovering around $\frac{N_\mathrm{dust}}{N_\mathrm{H}} \leq 10^{-12}$. The rise in erosion efficiency is conspicuous when looking at the mass distribution at the end of the collapse, where not only the tail of the distribution is affected but the distribution as a whole. Consequently, the eroding collisions between the small grains and the largest ones tend to lower the dust peak size at high density. To recapitulate, under standard assumptions (fiducial values for the other parameters) the ambipolar resistivity, $\eta_\mathrm{AD}$, is only slightly reduced prior to $n_\mathrm{H} = 10^{11} \ \mathrm{cm^{-3}}$ because the collisions involve grains with three orders of magnitude difference in size at most (collisions between nanometer and micrometer grains). At high density however, the dust distribution spans a much larger range of sizes and thus erosion is much more efficient (the said efficiency increases with the size difference between colliding grains). Consequently, the ambipolar and ohmic resistivities sharply rise, reaching values similar to the MRN case. The Hall resistivity is significantly affected for all densities $n_\mathrm{H} > 10^{8} \ \mathrm{cm^{-3}}$. Again, in this case, the excess of small grains enabled by erosion processes shifts the change of sign to higher densities. We stress that at high densities, grain-grain erosion significantly reduces the departure from the MRN reference in terms of magnetic resistivities.

\subsubsection{Impact of the ambipolar drift intensity, $\delta_\mathrm{AD}$}
\label{result impact of amibpolar drift}

  We now turn our attention to the impact of the ambipolar drift intensity $\delta_\mathrm{AD}$, which controls the amplitude of the collisional velocities induced by ambipolar diffusion (ambipolar drift velocity), as suggested in Eq. (\ref{AD efficiency}). In particular, we focus on its combined effect with grain-grain erosion and electrostatic repulsion.
Figure \ref{res_electrostatic_repulsion+delta} depicts the ambipolar resistivity $\eta_\mathrm{AD}$ evolution during the protostellar collapse with and without electrostatic repulsion between dust grains, with and without grain-grain erosion for different values of the ambipolar drift intensity. The second row conveys the same information but with a lower grain-grain sticking efficiency ($\epsilon_\mathrm{strick} = 0.5$). The same plot but for the ohmic resistivity can be found in the appendix (see Fig. \ref{resOHM_electrostatic_repulsion+delta}). 

First of all, by looking at the solid lines of the top left panel, we clearly see that reducing the ambipolar drift between grains alone leads to lower collisional velocities and thus to a much slower coagulation of the small grains. Their number drops less dramatically and consequently ambipolar resistivity $\eta_\mathrm{AD}$ is reduced at low density. Then, we turn on electrostatic repulsion (dotted lines) and remark that it has no impact whatsoever on the dust evolution when considering the reference values for the parameters (i.e. $\delta_\mathrm{AD} = 1$) because while Brownian motion is incapable of generating sufficiently large collisional velocities, the electrostatic barrier is easily overcome by the ambipolar diffusion drift (which allows for small grains to be collected by larger grains). However in reducing the ambipolar drift intensity to $\delta_\mathrm{AD} = 0.1$, we allowed the grains to experience collisional velocities lower than the Coulomb velocity threshold, thus inhibiting grain coagulation at low and intermediate densities, as suggested by Fig. \ref{res_electrostatic_repulsion+delta}. Indeed, with this scenario, the difference between the resulting ambipolar resistivity $\eta_\mathrm{AD}$ and that associated with an MRN dust distribution is of one order of magnitude at most. 
 Then, as the collapse proceeds and the density increases, the peak size of the dust distribution increases as well and ultimately the largest grains manage to collect the smallest ones via ambipolar drift even for low values of $\delta_\mathrm{AD}$, and the three dotted lines end up overlapping. This is due to the fact that the larger the difference in size (or equivalently in Hall factor), the higher the relative collisional velocity induced by ambipolar diffusion (ambipolar drift velocity), see Eq. (\ref{AD relative velocity}).
 
 In addition, we see on the top right panel that for $\delta_\mathrm{AD} = 1$, erosion is influential at high density only, as previously highlighted in Sect. \ref{erosion result section}. However, for $\delta_\mathrm{AD} = 10$, while electrostatic repulsion has no impact, grain-grain erosion is rendered very efficient even at low densities due to the large collisional velocities at play. Thus, the significant subsequent replenishment of small grains produces an ambipolar resistivity $\eta_\mathrm{AD}$ remarkably lower, close to the MRN case. Those two effects (electrostatic repulsion and erosion) can operate jointly at the same time but respectively allow us to reduce the ambipolar resistivity in different circumstances and environments.
 
 Finally, for $\delta_\mathrm{AD} = 1$, a look at the second row shows that a lower sticking efficiency slows down the coagulation of small grains (see Appendix \ref{res_sticking_efficiency} for a more detailed impact of the sticking efficiency). For $\delta_\mathrm{AD} = 0.1$, the combined effect of electrostatic repulsion and a lower sticking efficiency yields an even more evident reduction of the ambipolar resistivity. For $\delta_\mathrm{AD} = 10$, the very strong erosion is almost not affected by the reduced sticking efficiency. However, note that the replenishment of small grains induced by erosion at high densities in the case of $\delta_\mathrm{AD} = 0.1$ and  $\delta_\mathrm{AD} = 1$ is suppressed since half of the collisions lead to bouncing instead of erosion.

 In summary, on one hand, if the ambipolar drift is sufficiently weak, erosion has no influence but electrostatic repulsion inhibits small grains depletion as they repel each other instead of sticking. On the other hand, if the ambipolar drift is sufficiently strong, the Coulomb barrier is easily overcome but grain-grain erosion is efficient enough to repopulate the small grains and consequently lower the ambipolar resistivity $\eta_\mathrm{AD}$. We note that Appendix \ref{delta in PPD} explores the expected values of $\delta_\mathrm{AD}$ within a collapsing core.

\section{Dust growth: Potential consequences for the magnetic flux }
\label{magnetic field section}

\begin{figure}

\includegraphics[width= 0.5\textwidth]{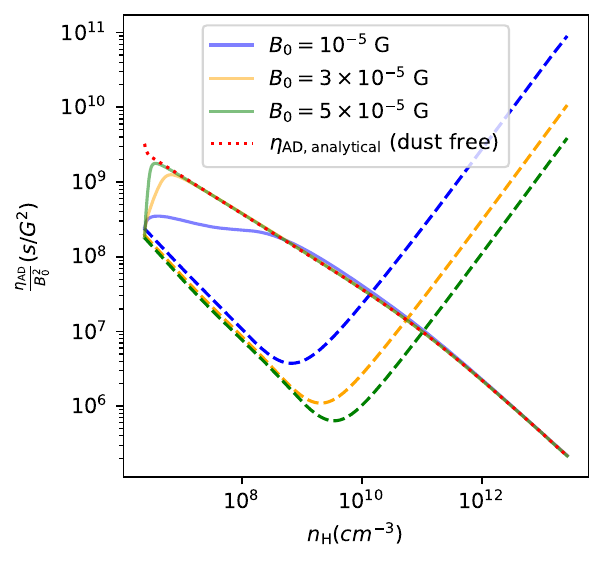}
\caption{Ambipolar resistivity, $\eta_\mathrm{AD}$, (normalized with the squared initial magnetic field strength) evolution with gas density throughout the protostellar collapse, for different values of the initial magnetic field strength. Solid lines: growing dust distribution ($\gamma_\mathrm{grain} = 190 \ \mathrm{erg \ cm^{-2}}$). Dashed lines : non evolving MRN dust distribution. Dotted red line is the analytical dust-free solution.} \label{eta ad for different B0}
\end{figure}

\begin{figure}

\includegraphics[width= 0.5\textwidth]{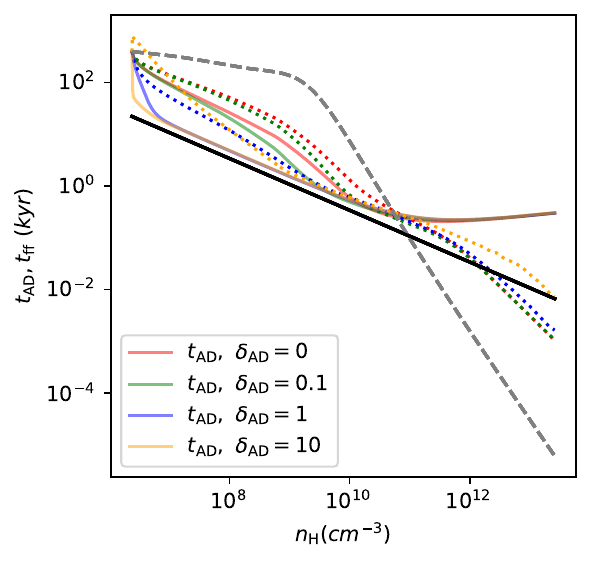}
\caption{Freefall timescale (in black) and ambipolar diffusion timescale as a function of gas density throughout the protostellar collapse for different values of the ambipolar drift intensity. The dashed grey line represents the fixed MRN case. The solid lines correspond to resistant grains ($\gamma_\mathrm{grain} = 190 \ \mathrm{erg \ cm^{-2}}$) whereas the dotted lines correspond to the fragile grains ($\gamma_\mathrm{grain} = 10 \ \mathrm{erg \ cm^{-2}}$).} \label{t_ad for different delta}
\end{figure}

\begin{figure}

\includegraphics[width= 0.5\textwidth]{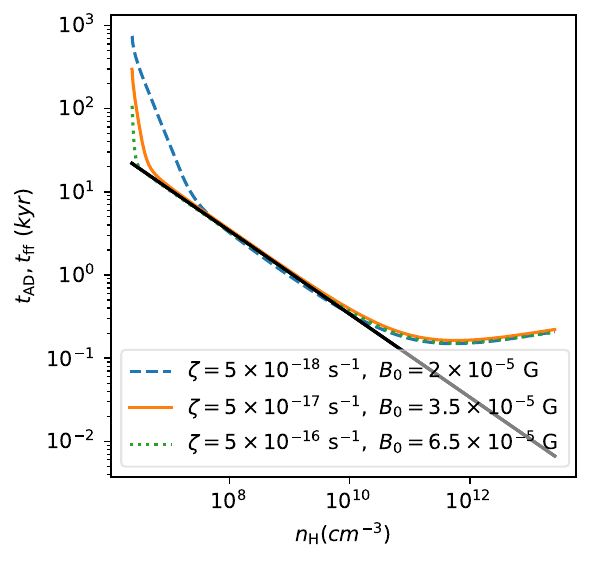}
\caption{Freefall timescale (in black) and ambipolar diffusion timescale as a function of gas density throughout the protostellar collapse for different values of the ionisation rate induced by cosmic rays. The initial magnetic field chosen corresponds to the critical value (minimum value) above which the small grain depletion leads to an ambipolar diffusion timescale lower than the freefall timescale.} \label{t_ad et B0 critical for different zeta}
\end{figure}

Assuming standard conditions (reference values for the parameters), but varying the initial magnetic field $B_0$ (see Eq. \ref{mag field equation}) and the ambipolar drift intensity $\delta_\mathrm{AD}$, we can focus on the evolution of the ambipolar resistivity, $\eta_\mathrm{AD}$, and on the ambipolar diffusion timescale, $t_\mathrm{AD}$, (see Eq. \ref{AD timescale}) during the collapse.

The results of Sect. \ref{Section results} highlight a tendency  for the magnetic resistivities to be significantly different from those one gets with a fixed MRN distribution. In particular, the ambipolar resistivity $\eta_\mathrm{AD}$ exhibits high values at low and intermediate densities due to small grain depletion enabled by ambipolar drift. Indeed, since the small grains contribute directly to the perpendicular conductivity at low density (see Sect. \ref{What species control the conductivities ?}), their removal leads to a sharp increase in the ambipolar resistivity. At high density however, the perpendicular conductivity is controlled by the ionic contribution. In the absence of small grains, ions, and electrons do not efficiently recombine and thus the ion density remains high. Therefore at high density, the ambipolar resistivity, $\eta_\mathrm{AD}$, is much lower with respect to the MRN case. The Ohm resistivity remains lower throughout the collapse, and monotonically increases. Unlike the fixed MRN case, the Hall resistivity systematically switches sign at some point, the corresponding density at which this transition occurs being highly sensitive to the dust distribution. 

\subsection{Small grain depletion: Grains cease to be the main charge carriers.}

Placing ourselves under standard assumptions, Fig. \ref{eta ad for different B0} depicts the evolution of the ambipolar resistivity $\eta_\mathrm{AD}$ during the collapse in the case of resilient grains ($\gamma_\mathrm{grain} = 190 \ \mathrm{erg \ cm^{-2}}$) for different values of the initial magnetic field strength $B_\mathrm{0}$. The dashed lines represent the fixed MRN reference case. The resistivity is normalized with $B_\mathrm{0}^2$ in such a way as to remove the direct dependency in $B^2$ and isolate and focus on the influence of the ambipolar diffusion on the resistivity. We wish to see to what extent the $\eta_\mathrm{AD} \propto B^2$ trend is relevant in the case of an evolving dust distribution. On Fig. \ref{eta ad for different B0}, we can see that for different values of the initial magnetic field strength, the normalized ambipolar resistivities $\frac{\eta_\mathrm{AD}}{B_\mathrm{0}^2}$ rapidly converge and overlap once the small grains have been removed, which hints at a $B^{2}$ dependency given by the analytical approximation, $\eta_\mathrm{AD} \propto \frac{B^2}{\gamma_\mathrm{in} \rho_\mathrm{i} \rho_\mathrm{n}}$ \citep{Duffin2008}, with $\gamma_\mathrm{in}$ as the drag coefficient between ions and neutrals, then $\rho_\mathrm{i}$ and $\rho_\mathrm{n}$ as the mass density of the ion and neutral fluids. This analytical expression, which is derived in considering only electrons and ions as charged species, holds true here because the small grains (which contributes to the ambipolar resistivity) are rapidly depleted by the ambipolar diffusion drift. Indeed, this is depicted by the sharp rise in the ambipolar resistivity, $\eta_\mathrm{AD}$, at the beginning of the collapse, the removal of small grains being even faster for a higher initial magnetic field strength (green  curve). Considering a balance between ionisation by cosmic-rays and ion/electron recombination, we have 
\begin{equation} \label{ion density}
n_\mathrm{ion} = \sqrt{\frac{\zeta n_\mathrm{H}}{\langle {\sigma v \rangle}}_\mathrm{ie}},   
\end{equation}
 for the numerical ion density when dust is negligible. Here, $\zeta$ is the ionisation rate induced by cosmic rays and ${\langle \sigma v \rangle}_\mathrm{ie} = 2 \times 10^{-7} \left (\frac{T}{300 \ \mathrm{K}} \right)^{-0.5} \ \mathrm{cm^{-3} \ s^{-1}}$ is the ion/electron recombination rate determined in \cite{Marchand2021}. To confirm this $B^2$ trend, we use directly the expression of the ambipolar resistivity $\eta_\mathrm{AD}$ given by Eq. (\ref{resistivities expressions}) to derive a complete expression. The parallel conductivity is dominated by the electron contribution. In absence of dust, the perpendicular one is dominated by the ion and the Hall conductivity is close to zero. Given that $\sigma_\mathrm{electron} \gg \sigma_\mathrm{ion,perp}$ and $\Gamma_\mathrm{ion} \gg 1$ we find that once the small grains have been removed, the ambipolar resistivity scales as 
\begin{equation}   
\eta_\mathrm{AD} \simeq 1/\sigma_{ion,perp} = \tau_\mathrm{ion} B^2/n_\mathrm{ion} c^2 \mu_\mathrm{ion} m_\mathrm{H} \propto 1/\sqrt{n_\mathrm{H}},
\end{equation}
 where we considered $B \propto \sqrt{n_\mathrm{H}}$ to switch from $B$ to $n_\mathrm{H}$, where $\sigma_\mathrm{ion}$ is the individual ion conductivity. $e$ is the electric charge, $c$ the speed of light in vacuum, and $\mu_\mathrm{ion} = 25$ (for $\mathrm{HCO^{+}}$ ions considered) and $\tau_\mathrm{ion}$ is the ion reduced temperature. $v_\mathrm{ion}$ is the thermal speed of the ions, and $k_\mathrm{B}$ is the Boltzmann constant. We recover the analytical form found in \cite{Duffin2008}. We overlay this analytical solution (dotted red curve) and see that it matches very well the numerical results, except of course at the first instants of the collapse, prior to small grain depletion.   
 
\subsection{A hint at a magnetic flux regulation}

Figure \ref{t_ad for different delta} pictures the evolution of the ambipolar diffusion timescale for different values of the ambipolar drift intensity $\delta_\mathrm{AD}$ (see Eq. \ref{AD efficiency}) in the case of resilient grains (solid lines) and fragile grains (dotted lines, $\gamma_\mathrm{grain} = 10 \ \mathrm{erg \ cm^{-2}}$). We can see on Fig. \ref{t_ad for different delta} that given the very high ambipolar resistivity $\eta_\mathrm{AD}$ at low density caused by the very swift depletion of small grains, the ambipolar diffusion timescale drops rapidly to a value very close to the free-fall timescale. As a consequence, the magnetic field should diffuse more efficiently. If $t_\mathrm{AD} > t_\mathrm{ff}$, the magnetic field strength should increase however less rapidly than the square root scaling of Eq. (\ref{mag field equation}). If $t_\mathrm{AD} = t_\mathrm{ff}$, it should dwell on its initial value since the increase in magnetic field by contraction of the flux tubes would be perfectly balanced by diffusion, or even decrease if $t_\mathrm{AD} < t_\mathrm{ff}$. It is only later on when the resistivity has sufficiently dropped that the magnetic field recouples to the gas and should increase with respect to the density.

The ambipolar drift can  be enhanced by either increasing the magnetic field strength or raising $\delta_\mathrm{AD}$. When fragmentation is inefficient ($\gamma_\mathrm{grain} = 190 \ \mathrm{erg \ cm^{-2}}$), we can see on Fig. \ref{t_ad for different delta} that in increasing the ambipolar drift intensity, the small grains are more swiftly depleted and thus the ambipolar diffusion timescale drops more rapidly. In addition, Fig. \ref{eta ad for different B0} shows that a higher magnetic field allows for the ambipolar resistivity, $\eta_\mathrm{AD}$, to rise more rapidly, which entails an ambipolar diffusion timescale dropping faster. Therefore we can identify here a mechanism of magnetic flux regulation. The higher the magnetic field, the stronger the ambipolar drift, the faster the small grains are removed and thus the faster the magnetic field diffuses. However when considering an efficient fragmentation ($\gamma_\mathrm{grain} = 10 \ \mathrm{erg \ cm^{-2}}$, dotted lines), the trend is the same except for high values of the ambipolar diffusion intensity. Indeed, in going from $\delta_\mathrm{AD} = 1$ to $\delta_\mathrm{AD} = 10$, it takes more time for the ambipolar diffusion timescale to drop, thus suppressing the magnetic flux regulation mechanism. This is due to the fact that for such fragile grains and high $\delta_\mathrm{AD}$, the ambipolar diffusion drift between dust grains induces destructive collisions leading to a replenishment of the small grains instead of a depletion of the latter. 

\subsection{Saturation of the magnetic field}

Finally, Fig. \ref{t_ad et B0 critical for different zeta} displays the evolution of the ambipolar diffusion timescale for different values of the initial magnetic field strength and ionisation rate induced by cosmic rays. The black line represents the free-fall timescale. 
The ambipolar diffusion timescale is computed via Eq. (\ref{AD timescale}) where each gas density is associated with a radius, $r,$ assuming a singular isothermal sphere \citep[see][]{Shu1977}
\begin{equation} \label{SIS}
    \rho = \frac{c_\mathrm{s}^2}{2 \pi G r^2}.   
\end{equation}
We can see that for a given value of the ionisation rate $\zeta$, there exists a critical value of the initial magnetic field for which the ambipolar diffusion timescale will inexorably drop to equate the freefall timescale. When it does so, the magnetic field diffuses in such a way as to remain constant. The larger the cosmic-rays ionisation rate, the better is the coupling between the plasma and the magnetic field (i.e. the lower the ambipolar resistivity, $\eta_\mathrm{AD}$). Consequently a higher initial magnetic field $B_\mathrm{0}$ is needed to produce an ambipolar drift strong enough to remove the small grains and, thus, to allow the magnetic field to diffuse significantly (i.e. to ensure that $t_\mathrm{AD} = t_\mathrm{ff}$).

 In such circumstances, we can define a saturation magnetic field that cannot be exceeded. We start from $t_\mathrm{AD} = t_\mathrm{ff}$ and assume that the ambipolar resistivity $\eta_\mathrm{AD}$ is controlled by the ion contribution, just as we did before. Using the analytical expression for $\eta_\mathrm{AD}$ which is plotted on Fig. \ref{eta ad for different B0}, the previously defined expression for the ion reduced temperature, the free-fall timescale defined in Eq. (\ref{frefall timescale}) and Eq. (\ref{ion density}), we get:\begin{equation}
\label{saturation B}
    B_\mathrm{sat}^2 = \frac{2 c_\mathrm{s}^2 \mu_\mathrm{ion}}{C(T)} \sqrt{\frac{\zeta 32 m_\mathrm{H}}{3 \pi {\langle \sigma v \rangle}_\mathrm{ie} \mu_\mathrm{g} \mathcal{G}}} n_\mathrm{H}.
\end{equation}
Whatever the initial magnetic field strength $B_\mathrm{0}$, the small grains will be removed, preventing the magnetic field strength from exceeding the value deduced from Eq. (\ref{saturation B}). We stress that this expression is valid as long as the dust contribution is absent, that is to say prior to a potential significant replenishment  of the small grains by fragmentation, and in the first instants of the collapse once they have been removed. We note, however, that at high densities in protoplanetary discs, the fragmentation barrier could be potentially reached after a few orbital periods, allowing for the small grain populations to be replenished and consequently allowing the gas and the magnetic field to decouple once more -- in regions where the temperature is below the sublimation temperature of the dust ($T \sim 1500 \ \mathrm{K}$) and where the ionisation induced by the protostar is negligible. Also, with rotation and magnetic pressure, the free-fall timescale should be longer and therefore diffusion of the magnetic flux would be even easier.

\section{Discussion} \label{Discussion}

\subsection{Considering whether there is a small grain depletion catastrophe}

In light of the soaring of the ambipolar resistivity, $\eta_\mathrm{AD}$, at low density resulting from dust coagulation, we may wonder whether small grain depletion has catastrophic consequences on protoplanetary disc formation. 

\subsubsection{Tensions with observed discs size}

Indeed, this paradigm is in tension with recent astronomical observations of protoplanetary discs, which appear to be rather small in size. Several surveys have been carried out, including CALYPSO \citep{Maury2019} for which less than 25 $\%$ of the Class 0 protostellar discs exhibit a radius $> 60 \ \mathrm{AU}$. Large discs seem to be rare and the observations favor the magnetized models of rotating protostellar collapse. This idea is supported  by pure hydrodynamical numerical simulations which yield a too high occurence of large discs \citep{Masson2016}. Even though turbulence effects can reduce the disc sizes without the need of a magnetic field  \citep[see][]{Bate2018}, the inclusion of a magnetic field produces much smaller discs \citep{Hennebelle2020,Lee2021}, more in agreement with observations. In their simulations of protoplanetary discs within massive star forming clumps, \cite{Lebreuilly2021} report a dominant population of small discs in both the ideal and non-ideal MHD (using an ambipolar resistivity obtained for a fixed MRN dust distribution) cases. For instance with ideal MHD, more than half of their disc population exhibit radii $< 50 \ \mathrm{AU}$ at the very early stages. When considering a higher ambipolar resistivity (via a truncated MRN dust distribution), \cite{Marchand2020} produce large discs while \cite{Zhao2016} suggest the possibility to form moderate protoplanetary disc sizes (a few tens of AU) when removing the smallest grains of the MRN dust distribution (grain sizes below $10 \ \mathrm{nm}$). Although different codes have been used, this discrepancy between both studies is most likely due to the $30^{\degree}$ angle set between the rotation axis of the core and the magnetic field in the first study as opposed to the perfect alignment in the second. Indeed, a misalignment between both axes tends to reduce magnetic breaking efficiency although not significantly in non-ideal MHD \citep{Masson2016}. Non-ideal effects are needed to prevent the magnetic braking catastrophe, but should be sufficiently weak in such a way as to produce disc sizes and outflows consistent with observations. 

\subsubsection{Ways to preserve the small grain population}

To alleviate the tension with observations, we identified several mechanisms preventing the small grains from being too severely depleted, thereby allowing us to reduce the ambipolar resistivity $\eta_\mathrm{AD}$ prior to $n_\mathrm{H} = 10^{10} \ \mathrm{cm^{-3}}$.

    Considering fragile grains: Fig. \ref{res_gamma_comparison_icenopermfrag} shows that values of $\gamma_{grain} \leq 10 \ \mathrm{erg \ cm^{-2}}$ yield a much lower ambipolar resistivity $\eta_\mathrm{AD}$ at intermediate densities as a consequence of the small grain replenishment induced by grain fragmentation. Since the elastic properties of the grains are quite uncertain, this scenario cannot be discarded.

    Low grain-grain sticking efficiency: Likewise the surface energy of the grains, this parameter remains poorly constrained. As depicted on Fig. \ref{res_sticking_efficiency}, allowing for one collision event out of ten to result in grain sticking allows us to recover an MRN-like ambipolar resistivity $\eta_\mathrm{AD}$ profile. In this case, the depletion of small grains is substantially curtailed.

    Strong or weak ambipolar drift: The ambipolar drift intensity $\delta_\mathrm{AD}$ appears as a prefactor in the ambipolar drift (Eq. (\ref{AD efficiency})) and embodies our ignorance of the phenomena. For instance, the current (the curl of the magnetic field) is approximated here using the Jeans length $\frac{|\vec{B}|}{\lambda_\mathrm{J}}$. This approximation is arguably simplistic and ignores directional effects inherent to a multi-dimensional problem. Indeed, it is well known that ambipolar diffusion is anisotropic and do not affect currents parallel to the magnetic field, as suggested by  the cross product in Eq. (\ref{AD vectorial form}).
    Ignoring here electrostatic repulsion and grain-grain erosion, Fig. \ref{res_electrostatic_repulsion+delta} exhibits a lower ambipolar resistivity $\eta_\mathrm{AD}$ when reducing $\delta_\mathrm{AD}$ by one order of magnitude. 

    Erosion: Even with resilient grains, erosion is capable of efficiently repopulating the small grains at low density provided that the ambipolar drift is strong enough ($\delta_\mathrm{AD} = 10$ on Fig. \ref{res_electrostatic_repulsion+delta}). Note that other destructive processes such as abrasion, gas-grain erosion, rotational disruption or grain sputtering could potentially enhance small grain replenishment as well.

    Electrostatic repulsion: With the reference value $\delta_\mathrm{AD} = 1$, the ambipolar drift generates grain collisional velocities larger than the velocity threshold imposed by the Coulomb barrier (see Eq. \ref{Coulomb barrier}). However, this is no longer the case when reducing the ambipolar drift intensity to $\delta_\mathrm{AD} = 0.1$. As a consequence, small grain depletion is further hindered and the combined effect of a lower ambipolar drift intensity and grain electrostatic repulsion leads to a significant reduction of the ambipolar resistivity $\eta_\mathrm{AD}$ (see Fig. \ref{res_electrostatic_repulsion+delta}). Similarly, keeping  $\delta_\mathrm{AD} = 1$, we notice that reducing the initial magnetic field generates a weaker ambipolar drift. Our simulations show that for an initial magnetic field strength (at $n_\mathrm{H} = 10^{4} \ \mathrm{cm^{-3}}$) of $B_0 \leq B_\mathrm{0,min} = 10 \ \mathrm{\mu G}$ (for $\zeta = 5 \times 10^{-17} \ \mathrm{s^{-1}}$ and $\delta_\mathrm{AD} = 1$, that is for an initial current strength of $\delta_\mathrm{AD}\frac{|\vec{B_\mathrm{0,min}}|}{\lambda_\mathrm{J}} = 5.72 \times 10^{-4} \ \mathrm{\mu G \ AU^{-1}}$), electrostatic repulsion effectively mitigates small grain coagulation and leads to a substantially lower ambipolar resistivity $\eta_\mathrm{AD}$. Equation (\ref{mag field equation}) is then a good approximation and the magnetic field can amplify in the early phase of the collapse. The value $B_\mathrm{0,min}$ below which electrostatic repulsion comes into play is somehow almost insensitive to the cosmic ionisation rate. We measured values between $9.5 \ \mathrm{\mu G}$ and $13.4 \ \mathrm{\mu G}$ for $\zeta$ in the range $\left [5 \times 10^{-18}, 5 \times 10^{-16} \right] \ \mathrm{s^{-1}}$. However, it should be kept in mind that although the population of small grains is preserved and the ambipolar resistivity is significantly reduced, a lower initial magnetic field (i.e. a higher mass-to-flux ratio) entails a weaker magnetic braking. 
    
Relying on electrostatic repulsion or erosion, we see from Fig. \ref{res_electrostatic_repulsion+delta} that in either scenario, the difference between the computed ambipolar resistivity value and that associated with an MRN dust distribution is of one order of magnitude at most (the ratio ranges between 2 at low density and 10 shortly before $n_\mathrm{H} = 10^{10} \ \mathrm{cm^{-3}}$). Relying on the analytical formulae for the radius of the early protoplanetary disc derived in \cite{Hennebelle2016} ($r_\mathrm{AD} \propto \eta_\mathrm{AD}^{\frac{2}{9}}$, and see \cite{Lee2024} for an extension to misalignment between magnetic field and rotation axis), we find that the corresponding difference in disc size would be of a factor of $10^{\frac{2}{9}} \sim 1.67$ in the worst case, in contrast to a factor of $100^{\frac{2}{9}} \sim 2.78$ without repulsion or erosion.  Comparing now the MRN case with the coagulation-repulsion case with a lower sticking efficiency ($\epsilon_\mathrm{stick} = 0.5$), we find a factor of $2^{\frac{2}{9}} \sim 1.16$. In terms of actual disc size values, we gather in Table \ref{table disc radius} the comparison between the different scenarios and readily see that we obtain reasonable disc radii with either repulsion or erosion.
The initial vertical magnetic field is assumed to be $B_\mathrm{z} = 30 \ \mathrm{\mu G}$, and the star/disc system mass to be $M = 0.1 \ M_{\odot}$ (consistent with a class 0 disc produced by the collapse of a one solar mass dense core). The ambipolar resistivity values $\eta_\mathrm{AD}$ are taken at the point of maximum departure from the MRN value, namely, slightly before $n_\mathrm{H} = 10^{10} \ \mathrm{cm^{-3}}$.

\begin{table}[]
\caption{disc radius $r_\mathrm{AD}$ for different scenarios.}

\centering
\begin{tabular}{ |p{5cm}||p{2cm}|}
 \hline
 \multicolumn{2}{|c|}{disc radius} \\
 \hline
 
 Fixed MRN&$30.4 \ \mathrm{AU}$\\

  Coagulation&$84.7 \ \mathrm{AU}$\\
  Coagulation-repulsion&$50.78 \ \mathrm{AU}$\\
  Coagulation-erosion&$50.78 \ \mathrm{AU}$\\
  Coagulation-erosion $\epsilon_\mathrm{stick} = 0.5$&$35.5 \ \mathrm{AU}$\\
\hline

\end{tabular}
\tablefoot{Disc radius $r_\mathrm{AD}$ \citep{Hennebelle2016} for different scenarios, with a fixed MRN dust distribution, accounting for coagulation only, coagulation and electrostatic repulsion, coagulation and erosion, coagulation-repulsion and a lower sticking efficiency ($\epsilon_\mathrm{stick} = 0.5$). The initial vertical magnetic field is assumed to be $B_\mathrm{z} = 30 \ \mathrm{\mu G}$, and the star-disc system mass to be $M = 0.1 \ M_{\odot}$. The ambipolar resistivity values $\eta_\mathrm{AD}$ are taken at the point of maximum departure from the MRN value, i.e. slightly before $n_\mathrm{H} = 10^{10} \ \mathrm{cm^{-3}}$.}.
\label{table disc radius}
\end{table}

For intermediate values of ambipolar drift intensity, $\delta_\mathrm{AD}$, neither electrostatic repulsion nor grain-grain erosion can significantly reduce the ambipolar resistivity. In such circumstances, we need to rely on other levers such as fragile grains or low sticking efficiency.
However,  the appendix offers a map of the ambipolar drift intensity $\delta_\mathrm{AD}$ in a collapsing core extracted from a 3D \texttt{RAMSES} \citep{Teyssier2002, Fromang2006} simulation performed by \cite{Lebreuilly2020}; namely, model mmMRNmhd at $t = 80 \ \mathrm{Kyr}$ (Fig. \ref{delta in PPD}). It is inferred by measuring the Lorentz force $\left(\vec{\nabla} \times \vec{B} \right) \times \vec{B}$ and comparing it with the approximation $\frac{|\vec{B}|^2}{\lambda_\mathrm{J}}$. As clearly seen, the intermediate value $\delta_\mathrm{AD} = 1$ is scarce, while both larger and lower values are ubiquitous. In the most central region (very young, nascent protoplanetary disc and core), and at large scale in the pseudo-disc, the ambipolar drift intensity dwells around $\delta_\mathrm{AD} = 10$. Consequently, grain-grain erosion is expected to induce a significant replenishment of the population of small grains in those denser regions. In contrast, the ambipolar drift intensity is much weaker ($\delta_\mathrm{AD} \sim 0.1$) in the rest of the domain, which hints at an inhibition of small grain coagulation via electrostatic repulsion in those diffuse regions. This reinforces the fundamental role that both mechanisms, electrostatic repulsion and grain-grain erosion, play in maintaining a reasonable ambipolar resistivity, $\eta_\mathrm{AD}$, during the protostellar collapse.

\subsection{Resilient grains or fragile grains:  Uncertainties in the collision outcomes}
\label{collision outcomes section}

Uncertainties remain in regard to the description of collision outcomes.
We recall here that some of them have been ignored in this work (see \cite{Wurm2021,Blum2018} for a detailed list). Bouncing for example, is expected to occur prior to fragmentation, and leads to compaction in place of grain sticking. The bouncing barrier has been widely described in the literature \citep{Zsom2010}. Erosion as a consequence of grain-gas drag could replenish the system with small grains but is expected to be relevant mainly in protoplanetary discs where dust grains can experience a strong gas head wind \citep{Rozner2020}. More relevant to our environment is grain-grain erosion \citep{Gundlach2015,Schrapler2018} whose influence has been investigated in this work. We based our model on the experimental results of \cite{Schrapler2018}.
In this regard, we found that grain-grain erosion is unconditionally very efficient at high density, where collisional velocities and size ratios are large, leading to noticeable variations in the Ohm and ambipolar resistivities. However, it is efficient at low density only for sufficiently strong ambipolar drifts.
Therefore, given the essential role of this process, future laboratory experiments will have to reinforce our understanding of this collision outcome and to further constrain the velocity threshold and efficiency. Additionally, electrostatic and magnetic reactions resulting from the charges carried by the grains may affect dust growth. In UV-shielded regions (where the grain photo-ionisation is negligible), the electrons are abundant ($n_\mathrm{e} > n_\mathrm{dust}$, i.e. efficient gas ionisation), plasma charging dominates,  and dust grains are expected to be negatively charged, on average \citep{Drain87}, electrostatic forces thus being repulsive and detrimental to grain coagulation \citep{Okuzumi2009,Akimkin2020}. When accounting for dust fragmentation and electrostatic repulsion, the population of sub-micron grains was found to be enhanced and the dust coagulation/fragmentation equilibrium to be severely affected (see \cite{Akimkin2023} who consider hot UV-shielded regions of protoplanetary discs where gas thermal ionisation is important, with $T = 1000 \ \mathrm{K}$ and $\rho = 10^{-12} \ \mathrm{g \ cm^{-3}}$). In the context of protostellar collapse, since the densities involved are very low, we believe that between two consecutive grain-grain collisions, the dust particles would have sufficient time to settle back to their average equilibrium negative charge value imposed by their interaction with electrons and ions. In this case, dust grains would feel repulsive electrostatic forces. As discussed previously, this can curtail to some extent the very efficient coagulation observed at the very beginning of the collapse.

Moreover, the conditions needed for sticking and fragmentation are diverse. Those relevant to a dense core environment are not well constrained in the literature either, and not well reproduced in laboratory experiments. That is why we explored a wide range of grain surface energies $\gamma_\mathrm{grain}$. For instance, collision outcomes for icy-grains have turned out to be temperature dependent \citep{Gartner2017}. Also, icy-grains have been believed to be  stickier that bare-silicate grains \citep{Gundlach2011,Drazkowska2017}, thus enhancing coagulation and being less prone to bouncing and compaction. Nevertheless, recent studies revealed that the grain surface energy $\gamma_\mathrm{grain}$ (which controls both the sticking efficiency and the fragmentation threshold) of icy-grains could sharply decrease from $T=200 \ \mathrm{K}$ to $T=175 \ \mathrm{K}$ due to a change in the crystal structure (again within the frame of the laboratory experiments, where $P=1 \ \mathrm{bar}$ and consequently the ice melting temperature being $T=273 \ \mathrm{K}$). Their measurements suggested a grain surface energy ranging between $\gamma_\mathrm{grain} = 2.9 \ \mathrm{erg \ cm^{-2}}$ at low temperature and $\gamma_\mathrm{grain} = 190 \ \mathrm{erg \ cm^{-2}}$ at higher temperature, which are the limit values we used in this work. For more details, refer to 
\cite{Musiolik2019}. Note however that amorphous ice at low pressure and temperature could recover an efficient sticking under UV irradiation \citep{Musiolik2021}. We explored the impact of the grain sticking efficiency in Sect. \ref{res_sticking_efficiency} and found that values above $0.3$ do not prevent the population of small grains from being depleted at low densities, thus allowing the ambipolar resistivity $\eta_\mathrm{AD}$ to dwell on very large values.

\subsection{Uncertainties in the turbulence model}
\label{uncertainties in turb model}

In this work, we rely on the model presented in \cite{Ormel2009} to compute the turbulent induced collisional velocities between dust grains.
This model assumes a Kolmogorov cascade of the turbulence, which is not necessarily expected in core collapse environments where magnetic and gravitational effects are likely to affect the nature of this energy cascade. In their recent work, \cite{Gong2021} expanded this model for different energy cascades. Moreover, there is a necessity to go beyond the approximations inherent to this model and to account for the back reaction of the dust onto the gas, which is not negligible in regions of high dust-to-gas ratio and at small scales. This backreaction would affect the behavior of the turbulence and has been showed to alter the propagation of Alfven waves \citep{Hennebelle2023}. Also, in order to better assess the turbulence induced collisional velocities, one needs to account for the magnetic field and the Lorentz force felt by the dust grains. Those refinements regarding the turbulence model would affect the growth of the peak size of the dust distribution and the replenishment of small grains via fragmentation of the largest ones.

In addition, turbulence can affect the local dust-to-gas ratio at small scales which is an essential ingredient in the context of dust growth. Indeed the turbulent coagulation timescale scales as the inverse of the dust-to-gas ratio squared. Turbulent motions are known to be capable of triggering hydrodynamics instabilities, allowing the dust-to-gas ratio to increase in certain regions \citep[Kelvin-Helmholtz instability for instance, see ][]{Hendrix2014}). Dust clumping by supersonic turbulence has been observed numerically in an environment reproducing molecular cloud physical conditions in \cite{Commerçon2023}, and in \cite{hopkins2016} for supersonic MHD turbulence. With both an analytical and numerical approach, \cite{Hopkins2018} expanded their own work on the Resonant Drag Instability (RDI) by including the role of magnetic fields and charged dust. As the dust and the gas drift relative to each other (e.g. via gravity), instabilities develop for all modes wavelength, leading to strong fluctuations in the dynamics of both fluids. In addition to the purely hydrodynamical instabilities observed in their previous work \citep{Hopkins2018a,Squire2018}, they identified a whole set of new instabilities. More generally, hydrodynamical instabilities represent promising mechanisms to enable the formation of planetesimals.

\section{Conclusion}
\label{Conclusion}

In this work, we study how the evolution of the dust distribution could affect the degree of coupling between the gas and the magnetic field of a collapsing core, including both the coagulation and fragmentation of dust grains. We also investigated the effect of grain-grain erosion, but we neglected other collision outcomes such as compaction, bouncing, and rotational disruption. We used the \texttt{shark} code to  self-consistently follow the evolution of the dust distribution and the subsequent impact on the magnetic resistivities. The code includes brownian motion, turbulence and ambipolar drift as grain-grain collision sources. We performed single zone simulations to save computational time and focus on the influence of fragmentation and various parameters. In summary, we found:

\begin{enumerate}

    \item As in previous studies \citep{Guillet2020,Kawasaki2022,Lebreuilly2023,Marchand2023}, dust growth is found to be significant and the subsequent impact on the magnetic resistivities very important. Therefore, dust evolution cannot be disregarded in numerical simulations.
    
    \item The degree of influence of the parameters depends on the fragmentation efficiency, which is primarily controlled by the grain surface energy. When considering a high value for this parameter ($\gamma_\mathrm{grain} = 190 \ \mathrm{erg \ cm^{-2}}$, i.e. resistant grains), fragmentation is substantially suppressed and the magnetic resistivities are only sensitive to the parameters which do not affect only (or not at all) the fragmentation processes; namely, the ambipolar drift intensity, the grain-grain erosion efficiency, electrostatic repulsion, and the cosmic-ray ionisation rate. However, when considering a low value for the grain surface energy ($\gamma_\mathrm{grain} = 10 \ \mathrm{erg \ cm^{-2}}$, fragile grains), fragmentation is severe and fragmentation-related parameters, such as the monomer size, the grain intrinsic density, and the power-law index of the fragment redistribution law become influential.

    \item  Grain fragmentation offers a way to replenish the population of small grains, which, in turn, significantly affects the resistivities. However, except for cases of very severe fragmentation, we find that the small grains are very rapidly removed due to ambipolar drift. As a consequence, the ohmic resistivity, $\eta_\mathrm{O}$, remains negligible with respect to the ambipolar resistivity, $\eta_\mathrm{AD}$, for the entirety of the collapse. In addition, the ambipolar resistivity is much higher at low densities and much lower at high densities than the value one gets in considering a fixed MRN dust distribution. Increasing the initial magnetic field strength leads to a more efficient depletion of small grains by ambipolar drift and a sharper rise in the ambipolar resistivity. The magnetic field then diffuses more rapidly, thus hinting at the regulation of the magnetic flux by ambipolar diffusion.
    
    \item To alleviate the tension between the remarkably high ambipolar resistivities obtained in our simulations and the statistically low protoplanetary disc sizes observed, we identified a few mechanisms that allow for  the ambipolar resistivity, $\eta_\mathrm{AD}$, to be reduced and, thus, to supposedly enhance magnetic braking: low grain surface energy (i.e. fragile grains), high ionisation rate via cosmic rays and low grain-grain sticking efficiency. In addition, we show that for a sufficiently low ambipolar drift on one hand, grain-grain electrostatic repulsion efficiently inhibits small grain depletion, leading to a much lower ambipolar resistivity prior of $n_\mathrm{H} = 10^{11} \ \mathrm{cm^{-3}}$ (whereas the grain-grain erosion is negligible). On the other hand, in the case of a strong ambipolar drift, the Coulomb barrier for electrostatic repulsion is easily overcome, however, grain-grain erosion causes a significant replenishment of the small grains; consequently, there is a sharp reduction of the ambipolar resistivity as a result. Both effects are expected to maintain a reasonable ambipolar resistivity, $\eta_\mathrm{AD}$, during the protostellar collapse.

    \item We pointed out the uncertainties regarding the elastic properties of the grains and the fragmentation model. Given the dramatic influence fragmentation can have on the magnetic resistivities, it is fundamental to better constrain the grain surface energy and the various collision outcomes relevant to interstellar dust in the adequate range of grain sizes. In addition, the grain surface energy also controls the maximum size dust grains can reach by the end of the collapse. For $\gamma_\mathrm{grain} = 10 \ \mathrm{erg \ cm^{-2}}$,  the grains grow up to $0.1 \ \mathrm{mm}$ while for $\gamma_\mathrm{grain} = 190 \ \mathrm{erg \ cm^{-2}}$, we manage to form grains of a few $\mathrm{mm}$.

\end{enumerate}

As a last note, we recall that we used the square root approximation of Eq. (\ref{mag field equation}) to compute the evolution of the magnetic field during the gravitational collapse as a function of the gas density only. However, the magnetic field affects the resistivities, which should (in turn) control the evolution of the latter through the induction equation and so forth. To rigorously describe the evolution of the magnetic field, multidimensional simulations, including dust growth along with a self-consistent computation of the magnetic field are needed.

%-------------------------------------------------------------------
 \begin{acknowledgements}

We thank the referee who helped us greatly in improving the clarity and relevancy of this paper. We thank the consortium and ERC (European Research Council) synergy grant ECOGAL (grant 855130) for their financial support, and members for their contribution through insightful ideas and remarks. 
 
\end{acknowledgements}

\bibliographystyle{aa}
\bibliography{ref}

\newcommand{\noop}[1]{}
\begin{thebibliography}{72}
\expandafter\ifx\csname natexlab\endcsname\relax\def\natexlab#1{#1}\fi

\bibitem[{{Akimkin} {et~al.}(2023){Akimkin}, {Ivlev}, {Caselli}, {Gong}, \& {Silsbee}}]{Akimkin2023}
{Akimkin}, V., {Ivlev}, A.~V., {Caselli}, P., {Gong}, M., \& {Silsbee}, K. 2023, \href{http://dx.doi.org/10.3847/1538-4357/ace2c5}{\color{magenta}\apj}, \href{https://ui.adsabs.harvard.edu/abs/2023ApJ...953...72A}{953, 72}

\bibitem[{{Akimkin} {et~al.}(2020){Akimkin}, {Ivlev}, \& {Caselli}}]{Akimkin2020}
{Akimkin}, V.~V., {Ivlev}, A.~V., \& {Caselli}, P. 2020, \href{http://dx.doi.org/10.3847/1538-4357/ab6299}{\color{magenta}\apj}, \href{https://ui.adsabs.harvard.edu/abs/2020ApJ...889...64A}{889, 64}

\bibitem[{{Bate}(2018)}]{Bate2018}
{Bate}, M.~R. 2018, \href{http://dx.doi.org/10.1093/mnras/sty169}{\color{magenta}\mnras}, \href{https://ui.adsabs.harvard.edu/abs/2018MNRAS.475.5618B}{475, 5618}

\bibitem[{{Bate}(2022)}]{Bate2022}
{Bate}, M.~R. 2022, \href{http://dx.doi.org/10.1093/mnras/stac1391}{\color{magenta}\mnras} \href{https://ui.adsabs.harvard.edu/abs/2022MNRAS.tmp.1351B}{[\eprint[arXiv]{2205.07681}]}

\bibitem[{{Blum}(2018)}]{Blum2018}
{Blum}, J. 2018, \href{http://dx.doi.org/10.1007/s11214-018-0486-5}{\color{magenta}\ssr}, \href{https://ui.adsabs.harvard.edu/abs/2018SSRv..214...52B}{214, 52}

\bibitem[{{Blum} \& {Wurm}(2000)}]{Blum2000}
{Blum}, J. \& {Wurm}, G. 2000, \href{http://dx.doi.org/10.1006/icar.1999.6234}{\color{magenta}\icarus}, \href{https://ui.adsabs.harvard.edu/abs/2000Icar..143..138B}{143, 138}

\bibitem[{{Commer{\c{c}}on} {et~al.}(2023){Commer{\c{c}}on}, {Lebreuilly}, {Price}, {Lovascio}, {Laibe}, \& {Hennebelle}}]{Commerçon2023}
{Commer{\c{c}}on}, B., {Lebreuilly}, U., {Price}, D.~J., {et~al.} 2023, \href{http://dx.doi.org/10.1051/0004-6361/202245141}{\color{magenta}\aap}, \href{https://ui.adsabs.harvard.edu/abs/2023A&A...671A.128C}{671, A128}

\bibitem[{{Dominik} \& {Tielens}(1997)}]{Dominik1997}
{Dominik}, C. \& {Tielens}, A.~G.~G.~M. 1997, \href{http://dx.doi.org/10.1086/303996}{\color{magenta}\apj}, \href{https://ui.adsabs.harvard.edu/abs/1997ApJ...480..647D}{480, 647}

\bibitem[{{Draine} \& {Sutin}(1987)}]{Drain87}
{Draine}, B.~T. \& {Sutin}, B. 1987, \href{http://dx.doi.org/10.1086/165596}{\color{magenta}\apj}, \href{https://ui.adsabs.harvard.edu/abs/1987ApJ...320..803D}{320, 803}

\bibitem[{{Drazkowska}(2017)}]{Drazkowska2017}
{Drazkowska}, J. 2017, in European Planetary Science Congress, \href{https://ui.adsabs.harvard.edu/abs/2017EPSC...11..815D}{EPSC2017--815}

\bibitem[{{Duffin} \& {Pudritz}(2008)}]{Duffin2008}
{Duffin}, D.~F. \& {Pudritz}, R.~E. 2008, \href{http://dx.doi.org/10.1111/j.1365-2966.2008.14026.x}{\color{magenta}\mnras}, \href{https://ui.adsabs.harvard.edu/abs/2008MNRAS.391.1659D}{391, 1659}

\bibitem[{{Epstein}(1924)}]{Epstein1924}
{Epstein}, P.~S. 1924, \href{http://dx.doi.org/10.1103/PhysRev.23.710}{\color{magenta}Physical Review}, \href{http://adsabs.harvard.edu/abs/1924PhRv...23..710E}{23, 710}

\bibitem[{{Fromang} {et~al.}(2006){Fromang}, {Hennebelle}, \& {Teyssier}}]{Fromang2006}
{Fromang}, S., {Hennebelle}, P., \& {Teyssier}, R. 2006, \href{http://dx.doi.org/10.1051/0004-6361:20065371}{\color{magenta}\aap}, \href{http://adsabs.harvard.edu/abs/2006A%26A...457..371F}{457, 371}

\bibitem[{{G{\"a}rtner} {et~al.}(2017){G{\"a}rtner}, {Gundlach}, {Headen}, {Ratte}, {Oesert}, {Gorb}, {Youngs}, {Bowron}, {Blum}, \& {Fraser}}]{Gartner2017}
{G{\"a}rtner}, S., {Gundlach}, B., {Headen}, T.~F., {et~al.} 2017, \href{http://dx.doi.org/10.3847/1538-4357/aa8c7f}{\color{magenta}\apj}, \href{https://ui.adsabs.harvard.edu/abs/2017ApJ...848...96G}{848, 96}

\bibitem[{{Gong} {et~al.}(2021){Gong}, {Ivlev}, {Akimkin}, \& {Caselli}}]{Gong2021}
{Gong}, M., {Ivlev}, A.~V., {Akimkin}, V., \& {Caselli}, P. 2021, \href{http://dx.doi.org/10.3847/1538-4357/ac0ce8}{\color{magenta}\apj}, \href{https://ui.adsabs.harvard.edu/abs/2021ApJ...917...82G}{917, 82}

\bibitem[{{Gould} \& {Salpeter}(1963)}]{Gould1963}
{Gould}, R.~J. \& {Salpeter}, E.~E. 1963, \href{http://dx.doi.org/10.1086/147654}{\color{magenta}\apj}, \href{https://ui.adsabs.harvard.edu/abs/1963ApJ...138..393G}{138, 393}

\bibitem[{{Guillet} {et~al.}(2020){Guillet}, {Hennebelle}, {Pineau des For{\^e}ts}, {Marcowith}, {Commer{\c{c}}on}, \& {Marchand}}]{Guillet2020}
{Guillet}, V., {Hennebelle}, P., {Pineau des For{\^e}ts}, G., {et~al.} 2020, \href{http://dx.doi.org/10.1051/0004-6361/201937387}{\color{magenta}\aap}, \href{https://ui.adsabs.harvard.edu/abs/2020A&A...643A..17G}{643, A17}

\bibitem[{{Guillet} {et~al.}(2007){Guillet}, {Pineau Des For{\^e}ts}, \& {Jones}}]{Guillet2007}
{Guillet}, V., {Pineau Des For{\^e}ts}, G., \& {Jones}, A.~P. 2007, \href{http://dx.doi.org/10.1051/0004-6361:20078094}{\color{magenta}\aap}, \href{https://ui.adsabs.harvard.edu/abs/2007A&A...476..263G}{476, 263}

\bibitem[{{Gundlach} \& {Blum}(2015)}]{Gundlach2015}
{Gundlach}, B. \& {Blum}, J. 2015, \href{http://dx.doi.org/10.1088/0004-637X/798/1/34}{\color{magenta}\apj}, \href{https://ui.adsabs.harvard.edu/abs/2015ApJ...798...34G}{798, 34}

\bibitem[{{Gundlach} {et~al.}(2011){Gundlach}, {Kilias}, {Beitz}, \& {Blum}}]{Gundlach2011}
{Gundlach}, B., {Kilias}, S., {Beitz}, E., \& {Blum}, J. 2011, \href{http://dx.doi.org/10.1016/j.icarus.2011.05.005}{\color{magenta}\icarus}, \href{https://ui.adsabs.harvard.edu/abs/2011Icar..214..717G}{214, 717}

\bibitem[{{Hendrix} \& {Keppens}(2014)}]{Hendrix2014}
{Hendrix}, T. \& {Keppens}, R. 2014, \href{http://dx.doi.org/10.1051/0004-6361/201322322}{\color{magenta}\aap}, \href{https://ui.adsabs.harvard.edu/abs/2014A&A...562A.114H}{562, A114}

\bibitem[{{Hennebelle} {et~al.}(2016){Hennebelle}, {Commer{\c{c}}on}, {Chabrier}, \& {Marchand}}]{Hennebelle2016}
{Hennebelle}, P., {Commer{\c{c}}on}, B., {Chabrier}, G., \& {Marchand}, P. 2016, \href{http://dx.doi.org/10.3847/2041-8205/830/1/L8}{\color{magenta}\apjl}, \href{https://ui.adsabs.harvard.edu/abs/2016ApJ...830L...8H}{830, L8}

\bibitem[{{Hennebelle} {et~al.}(2020){Hennebelle}, {Commer{\c{c}}on}, {Lee}, \& {Chabrier}}]{Hennebelle2020}
{Hennebelle}, P., {Commer{\c{c}}on}, B., {Lee}, Y.-N., \& {Chabrier}, G. 2020, \href{http://dx.doi.org/10.3847/1538-4357/abbfab}{\color{magenta}\apj}, \href{https://ui.adsabs.harvard.edu/abs/2020ApJ...904..194H}{904, 194}

\bibitem[{{Hennebelle} \& {Lebreuilly}(2023)}]{Hennebelle2023}
{Hennebelle}, P. \& {Lebreuilly}, U. 2023, \href{http://dx.doi.org/10.1051/0004-6361/202245120}{\color{magenta}\aap}, \href{https://ui.adsabs.harvard.edu/abs/2023A&A...674A.149H}{674, A149}

\bibitem[{{Hopkins} \& {Lee}(2016)}]{hopkins2016}
{Hopkins}, P.~F. \& {Lee}, H. 2016, \href{http://dx.doi.org/10.1093/mnras/stv2745}{\color{magenta}\mnras}, \href{https://ui.adsabs.harvard.edu/abs/2016MNRAS.456.4174H}{456, 4174}

\bibitem[{{Hopkins} \& {Squire}(2018{\natexlab{a}})}]{Hopkins2018a}
{Hopkins}, P.~F. \& {Squire}, J. 2018{\natexlab{a}}, \href{http://dx.doi.org/10.1093/mnras/sty1982}{\color{magenta}\mnras}, \href{https://ui.adsabs.harvard.edu/abs/2018MNRAS.480.2813H}{480, 2813}

\bibitem[{{Hopkins} \& {Squire}(2018{\natexlab{b}})}]{Hopkins2018}
{Hopkins}, P.~F. \& {Squire}, J. 2018{\natexlab{b}}, \href{http://dx.doi.org/10.1093/mnras/sty1604}{\color{magenta}\mnras}, \href{https://ui.adsabs.harvard.edu/abs/2018MNRAS.479.4681H}{479, 4681}

\bibitem[{{Kawasaki} {et~al.}(2022){Kawasaki}, {Koga}, \& {Machida}}]{Kawasaki2022}
{Kawasaki}, Y., {Koga}, S., \& {Machida}, M.~N. 2022, \href{http://dx.doi.org/10.1093/mnras/stac1919}{\color{magenta}\mnras}, \href{https://ui.adsabs.harvard.edu/abs/2022MNRAS.515.2072K}{515, 2072}

\bibitem[{{Kawasaki} \& {Machida}(2023)}]{Kawasaki2023}
{Kawasaki}, Y. \& {Machida}, M.~N. 2023, \href{http://dx.doi.org/10.1093/mnras/stad1241}{\color{magenta}\mnras} \href{https://ui.adsabs.harvard.edu/abs/2023MNRAS.tmp.1177K}{[\eprint[arXiv]{2304.13271}]}

\bibitem[{{Lebreuilly} {et~al.}(2020){Lebreuilly}, {Commer{\c{c}}on}, \& {Laibe}}]{Lebreuilly2020}
{Lebreuilly}, U., {Commer{\c{c}}on}, B., \& {Laibe}, G. 2020, \href{http://dx.doi.org/10.1051/0004-6361/202038174}{\color{magenta}\aap}, \href{https://ui.adsabs.harvard.edu/abs/2020A&A...641A.112L}{641, A112}

\bibitem[{{Lebreuilly} {et~al.}(2021){Lebreuilly}, {Hennebelle}, {Colman}, {Commer{\c{c}}on}, {Klessen}, {Maury}, {Molinari}, \& {Testi}}]{Lebreuilly2021}
{Lebreuilly}, U., {Hennebelle}, P., {Colman}, T., {et~al.} 2021, \href{http://dx.doi.org/10.3847/2041-8213/ac158c}{\color{magenta}\apjl}, \href{https://ui.adsabs.harvard.edu/abs/2021ApJ...917L..10L}{917, L10}

\bibitem[{{Lebreuilly} {et~al.}(2023){Lebreuilly}, {Vallucci-Goy}, {Guillet}, {Lombart}, \& {Marchand}}]{Lebreuilly2023}
{Lebreuilly}, U., {Vallucci-Goy}, V., {Guillet}, V., {Lombart}, M., \& {Marchand}, P. 2023, \href{http://dx.doi.org/10.1093/mnras/stac3220}{\color{magenta}\mnras}, \href{https://ui.adsabs.harvard.edu/abs/2023MNRAS.518.3326L}{518, 3326}

\bibitem[{{Lee} {et~al.}(2021){Lee}, {Marchand}, {Liu}, \& {Hennebelle}}]{Lee2021}
{Lee}, Y.-N., {Marchand}, P., {Liu}, Y.-H., \& {Hennebelle}, P. 2021, \href{http://dx.doi.org/10.3847/1538-4357/ac235d}{\color{magenta}\apj}, \href{https://ui.adsabs.harvard.edu/abs/2021ApJ...922...36L}{922, 36}

\bibitem[{{Lee} {et~al.}(2024){Lee}, {Ray}, {Marchand}, \& {Hennebelle}}]{Lee2024}
{Lee}, Y.-N., {Ray}, B., {Marchand}, P., \& {Hennebelle}, P. 2024, \href{http://dx.doi.org/10.3847/2041-8213/ad192a}{\color{magenta}\apjl}, \href{https://ui.adsabs.harvard.edu/abs/2024ApJ...961L..28L}{961, L28}

\bibitem[{{Li} {et~al.}(2011){Li}, {Krasnopolsky}, \& {Shang}}]{Li2011}
{Li}, Z.-Y., {Krasnopolsky}, R., \& {Shang}, H. 2011, \href{http://dx.doi.org/10.1088/0004-637X/738/2/180}{\color{magenta}\apj}, \href{https://ui.adsabs.harvard.edu/abs/2011ApJ...738..180L}{738, 180}

\bibitem[{{Machida} {et~al.}(2006){Machida}, {Inutsuka}, \& {Matsumoto}}]{Machida2006}
{Machida}, M.~N., {Inutsuka}, S.-i., \& {Matsumoto}, T. 2006, \href{http://dx.doi.org/10.1086/507179}{\color{magenta}\apjl}, \href{https://ui.adsabs.harvard.edu/abs/2006ApJ...647L.151M}{647, L151}

\bibitem[{{Marchand} {et~al.}(2021){Marchand}, {Guillet}, {Lebreuilly}, \& {Mac Low}}]{Marchand2021}
{Marchand}, P., {Guillet}, V., {Lebreuilly}, U., \& {Mac Low}, M.~M. 2021, \href{http://dx.doi.org/10.1051/0004-6361/202040077}{\color{magenta}\aap}, \href{https://ui.adsabs.harvard.edu/abs/2021A&A...649A..50M}{649, A50}

\bibitem[{{Marchand} {et~al.}(2023){Marchand}, {Lebreuilly}, {Mac Low}, \& {Guillet}}]{Marchand2023}
{Marchand}, P., {Lebreuilly}, U., {Mac Low}, M.~M., \& {Guillet}, V. 2023, \href{http://dx.doi.org/10.1051/0004-6361/202244291}{\color{magenta}\aap}, \href{https://ui.adsabs.harvard.edu/abs/2023A&A...670A..61M}{670, A61}

\bibitem[{{Marchand} {et~al.}(2016){Marchand}, {Masson}, {Chabrier}, {Hennebelle}, {Commer{\c c}on}, \& {Vaytet}}]{Marchand2016}
{Marchand}, P., {Masson}, J., {Chabrier}, G., {et~al.} 2016, \href{http://dx.doi.org/10.1051/0004-6361/201526780}{\color{magenta}\aap}, \href{http://adsabs.harvard.edu/abs/2016A%26A...592A..18M}{592, A18}

\bibitem[{{Marchand} {et~al.}(2020){Marchand}, {Tomida}, {Tanaka}, {Commer{\c{c}}on}, \& {Chabrier}}]{Marchand2020}
{Marchand}, P., {Tomida}, K., {Tanaka}, K. E.~I., {Commer{\c{c}}on}, B., \& {Chabrier}, G. 2020, \href{http://dx.doi.org/10.3847/1538-4357/abad99}{\color{magenta}\apj}, \href{https://ui.adsabs.harvard.edu/abs/2020ApJ...900..180M}{900, 180}

\bibitem[{{Masson} {et~al.}(2016){Masson}, {Chabrier}, {Hennebelle}, {Vaytet}, \& {Commer{\c{c}}on}}]{Masson2016}
{Masson}, J., {Chabrier}, G., {Hennebelle}, P., {Vaytet}, N., \& {Commer{\c{c}}on}, B. 2016, \href{http://dx.doi.org/10.1051/0004-6361/201526371}{\color{magenta}\aap}, \href{https://ui.adsabs.harvard.edu/abs/2016A&A...587A..32M}{587, A32}

\bibitem[{{Mathis} {et~al.}(1977){Mathis}, {Rumpl}, \& {Nordsieck}}]{Mathis1977}
{Mathis}, J.~S., {Rumpl}, W., \& {Nordsieck}, K.~H. 1977, \href{http://dx.doi.org/10.1086/155591}{\color{magenta}\apj}, \href{http://adsabs.harvard.edu/abs/1977ApJ...217..425M}{217, 425}

\bibitem[{{Maury} {et~al.}(2019){Maury}, {Andr{\'e}}, {Testi}, {Maret}, {Belloche}, {Hennebelle}, {Cabrit}, {Codella}, {Gueth}, {Podio}, {Anderl}, {Bacmann}, {Bontemps}, {Gaudel}, {Ladjelate}, {Lef{\`e}vre}, {Tabone}, \& {Lefloch}}]{Maury2019}
{Maury}, A.~J., {Andr{\'e}}, P., {Testi}, L., {et~al.} 2019, \href{http://dx.doi.org/10.1051/0004-6361/201833537}{\color{magenta}\aap}, \href{https://ui.adsabs.harvard.edu/abs/2019A&A...621A..76M}{621, A76}

\bibitem[{{McKee} \& {Ostriker}(2007)}]{McKee2007}
{McKee}, C.~F. \& {Ostriker}, E.~C. 2007, \href{http://dx.doi.org/10.1146/annurev.astro.45.051806.110602}{\color{magenta}\araa}, \href{https://ui.adsabs.harvard.edu/abs/2007ARA&A..45..565M}{45, 565}

\bibitem[{{Mouschovias}(1991)}]{Mouschovias1991}
{Mouschovias}, T.~C. 1991, \href{http://dx.doi.org/10.1086/170035}{\color{magenta}\apj}, \href{https://ui.adsabs.harvard.edu/abs/1991ApJ...373..169M}{373, 169}

\bibitem[{{Musiolik}(2021)}]{Musiolik2021}
{Musiolik}, G. 2021, \href{http://dx.doi.org/10.1093/mnras/stab1963}{\color{magenta}\mnras}, \href{https://ui.adsabs.harvard.edu/abs/2021MNRAS.506.5153M}{506, 5153}

\bibitem[{{Musiolik} \& {Wurm}(2019)}]{Musiolik2019}
{Musiolik}, G. \& {Wurm}, G. 2019, \href{http://dx.doi.org/10.3847/1538-4357/ab0428}{\color{magenta}\apj}, \href{https://ui.adsabs.harvard.edu/abs/2019ApJ...873...58M}{873, 58}

\bibitem[{{Nakano} {et~al.}(2002){Nakano}, {Nishi}, \& {Umebayashi}}]{Nakano2002}
{Nakano}, T., {Nishi}, R., \& {Umebayashi}, T. 2002, \href{http://dx.doi.org/10.1086/340587}{\color{magenta}\apj}, \href{https://ui.adsabs.harvard.edu/abs/2002ApJ...573..199N}{573, 199}

\bibitem[{{Okuzumi}(2009)}]{Okuzumi2009}
{Okuzumi}, S. 2009, \href{http://dx.doi.org/10.1088/0004-637X/698/2/1122}{\color{magenta}\apj}, \href{https://ui.adsabs.harvard.edu/abs/2009ApJ...698.1122O}{698, 1122}

\bibitem[{{Ormel} \& {Cuzzi}(2007)}]{Ormel2007}
{Ormel}, C.~W. \& {Cuzzi}, J.~N. 2007, \href{http://dx.doi.org/10.1051/0004-6361:20066899}{\color{magenta}\aap}, \href{https://ui.adsabs.harvard.edu/abs/2007A&A...466..413O}{466, 413}

\bibitem[{{Ormel} {et~al.}(2009){Ormel}, {Paszun}, {Dominik}, \& {Tielens}}]{Ormel2009}
{Ormel}, C.~W., {Paszun}, D., {Dominik}, C., \& {Tielens}, A.~G.~G.~M. 2009, \href{http://dx.doi.org/10.1051/0004-6361/200811158}{\color{magenta}\aap}, \href{https://ui.adsabs.harvard.edu/abs/2009A&A...502..845O}{502, 845}

\bibitem[{{Padovani} {et~al.}(2022){Padovani}, {Bialy}, {Galli}, {Ivlev}, {Grassi}, {Scarlett}, {Rehill}, {Zammit}, {Fursa}, \& {Bray}}]{Padovani2022}
{Padovani}, M., {Bialy}, S., {Galli}, D., {et~al.} 2022, \href{http://dx.doi.org/10.1051/0004-6361/202142560}{\color{magenta}\aap}, \href{https://ui.adsabs.harvard.edu/abs/2022A&A...658A.189P}{658, A189}

\bibitem[{{Rozner} {et~al.}(2020){Rozner}, {Grishin}, \& {Perets}}]{Rozner2020}
{Rozner}, M., {Grishin}, E., \& {Perets}, H.~B. 2020, \href{http://dx.doi.org/10.1093/mnras/staa1864}{\color{magenta}\mnras}, \href{https://ui.adsabs.harvard.edu/abs/2020MNRAS.496.4827R}{496, 4827}

\bibitem[{{Schrapler} {et~al.}(2018){Schrapler}, {Blum}, {Krijt}, \& {Raabe}}]{Schrapler2018}
{Schrapler}, R., {Blum}, J., {Krijt}, S., \& {Raabe}, J.-H. 2018, \href{http://dx.doi.org/10.3847/1538-4357/aaa0d2}{\color{magenta}\apj}, \href{https://ui.adsabs.harvard.edu/abs/2018ApJ...853...74S}{853, 74}

\bibitem[{{Shu}(1977)}]{Shu1977}
{Shu}, F.~H. 1977, \href{http://dx.doi.org/10.1086/155274}{\color{magenta}\apj}, \href{https://ui.adsabs.harvard.edu/abs/1977ApJ...214..488S}{214, 488}

\bibitem[{{Silsbee} {et~al.}(2020){Silsbee}, {Ivlev}, {Sipil{\"a}}, {Caselli}, \& {Zhao}}]{Silsbee2020}
{Silsbee}, K., {Ivlev}, A.~V., {Sipil{\"a}}, O., {Caselli}, P., \& {Zhao}, B. 2020, \href{http://dx.doi.org/10.1051/0004-6361/202038063}{\color{magenta}\aap}, \href{https://ui.adsabs.harvard.edu/abs/2020A&A...641A..39S}{641, A39}

\bibitem[{{Smoluchowski}(1916)}]{Smolu16}
{Smoluchowski}, M.~V. 1916, Zeitschrift fur Physik, \href{https://ui.adsabs.harvard.edu/abs/1916ZPhy...17..557S}{17, 557}

\bibitem[{{Spitzer}(1941)}]{Spitzer1941}
{Spitzer}, Lyman, J. 1941, \href{http://dx.doi.org/10.1086/144273}{\color{magenta}\apj}, \href{https://ui.adsabs.harvard.edu/abs/1941ApJ....93..369S}{93, 369}

\bibitem[{{Squire} \& {Hopkins}(2018)}]{Squire2018}
{Squire}, J. \& {Hopkins}, P.~F. 2018, \href{http://dx.doi.org/10.1093/mnras/sty854}{\color{magenta}\mnras}, \href{https://ui.adsabs.harvard.edu/abs/2018MNRAS.477.5011S}{477, 5011}

\bibitem[{{Testi} {et~al.}(2014){Testi}, {Birnstiel}, {Ricci}, {Andrews}, {Blum}, {Carpenter}, {Dominik}, {Isella}, {Natta}, {Williams}, \& {Wilner}}]{Testi2014}
{Testi}, L., {Birnstiel}, T., {Ricci}, L., {et~al.} 2014, in Protostars and Planets VI, ed. H.~{Beuther}, R.~S. {Klessen}, C.~P. {Dullemond}, \& T.~{Henning}, \href{https://ui.adsabs.harvard.edu/abs/2014prpl.conf..339T}{339}

\bibitem[{{Teyssier}(2002)}]{Teyssier2002}
{Teyssier}, R. 2002, \href{http://dx.doi.org/10.1051/0004-6361:20011817}{\color{magenta}\aap}, \href{http://adsabs.harvard.edu/abs/2002A%26A...385..337T}{385, 337}

\bibitem[{{Troland} \& {Crutcher}(2008)}]{Troland2008}
{Troland}, T.~H. \& {Crutcher}, R.~M. 2008, \href{http://dx.doi.org/10.1086/587546}{\color{magenta}\apj}, \href{https://ui.adsabs.harvard.edu/abs/2008ApJ...680..457T}{680, 457}

\bibitem[{{Tsukamoto} {et~al.}(2021){Tsukamoto}, {Machida}, \& {Inutsuka}}]{Tsukamoto2021}
{Tsukamoto}, Y., {Machida}, M.~N., \& {Inutsuka}, S.-i. 2021, \href{http://dx.doi.org/10.3847/2041-8213/ac2b2f}{\color{magenta}\apjl}, \href{https://ui.adsabs.harvard.edu/abs/2021ApJ...920L..35T}{920, L35}

\bibitem[{{Tsukamoto} {et~al.}(2023){Tsukamoto}, {Machida}, \& {Inutsuka}}]{Tsukamoto2023}
{Tsukamoto}, Y., {Machida}, M.~N., \& {Inutsuka}, S.-i. 2023, \href{http://dx.doi.org/10.1093/pasj/psad040}{\color{magenta}\pasj}, \href{https://ui.adsabs.harvard.edu/abs/2023PASJ...75..835T}{75, 835}

\bibitem[{{Tu} {et~al.}(2022){Tu}, {Li}, \& {Lam}}]{Tu2022}
{Tu}, Y., {Li}, Z.-Y., \& {Lam}, K.~H. 2022, \href{http://dx.doi.org/10.1093/mnras/stac2030}{\color{magenta}\mnras} \href{https://ui.adsabs.harvard.edu/abs/2022MNRAS.tmp.1903T}{[\eprint[arXiv]{2207.14151}]}

\bibitem[{{Voelk} {et~al.}(1980){Voelk}, {Jones}, {Morfill}, \& {Roeser}}]{voelk1980}
{Voelk}, H.~J., {Jones}, F.~C., {Morfill}, G.~E., \& {Roeser}, S. 1980, \aap, \href{https://ui.adsabs.harvard.edu/abs/1980A&A....85..316V}{85, 316}

\bibitem[{{Vorobyov} \& {Elbakyan}(2019)}]{Vorobyov2019}
{Vorobyov}, E.~I. \& {Elbakyan}, V.~G. 2019, \href{http://dx.doi.org/10.1051/0004-6361/201936132}{\color{magenta}\aap}, \href{https://ui.adsabs.harvard.edu/abs/2019A&A...631A...1V}{631, A1}

\bibitem[{{Wurm} \& {Teiser}(2021)}]{Wurm2021}
{Wurm}, G. \& {Teiser}, J. 2021, \href{http://dx.doi.org/10.1038/s42254-021-00312-7}{\color{magenta}Nature Reviews Physics}, \href{https://ui.adsabs.harvard.edu/abs/2021NatRP...3..405W}{3, 405}

\bibitem[{{Wurster} {et~al.}(2018){Wurster}, {Bate}, \& {Price}}]{Wurster2018}
{Wurster}, J., {Bate}, M.~R., \& {Price}, D.~J. 2018, \href{http://dx.doi.org/10.1093/mnras/stx3339}{\color{magenta}\mnras}, \href{https://ui.adsabs.harvard.edu/abs/2018MNRAS.475.1859W}{475, 1859}

\bibitem[{{Wurster} {et~al.}(2016){Wurster}, {Price}, \& {Bate}}]{Wurster2016}
{Wurster}, J., {Price}, D.~J., \& {Bate}, M.~R. 2016, \href{http://dx.doi.org/10.1093/mnras/stw013}{\color{magenta}\mnras}, \href{https://ui.adsabs.harvard.edu/abs/2016MNRAS.457.1037W}{457, 1037}

\bibitem[{{Zhao} {et~al.}(2016){Zhao}, {Caselli}, {Li}, {Krasnopolsky}, {Shang}, \& {Nakamura}}]{Zhao2016}
{Zhao}, B., {Caselli}, P., {Li}, Z.-Y., {et~al.} 2016, \href{http://dx.doi.org/10.1093/mnras/stw1124}{\color{magenta}\mnras}, \href{http://adsabs.harvard.edu/abs/2016MNRAS.460.2050Z}{460, 2050}

\bibitem[{{Zsom} {et~al.}(2010){Zsom}, {Ormel}, {Blum}, {Guettler}, \& {Dullemond}}]{Zsom2010}
{Zsom}, A., {Ormel}, C., {Blum}, J., {Guettler}, C., \& {Dullemond}, C. 2010, in EGU General Assembly Conference Abstracts, EGU General Assembly Conference Abstracts, \href{https://ui.adsabs.harvard.edu/abs/2010EGUGA..12.2966Z}{2966}

\end{thebibliography}

\appendix
\label{appendix}

\section{Which species controls the conductivities during collapse}
\label{What species control the conductivities ?}

\begin{figure*}[hbt]
    \includegraphics[width= 1.0\textwidth]{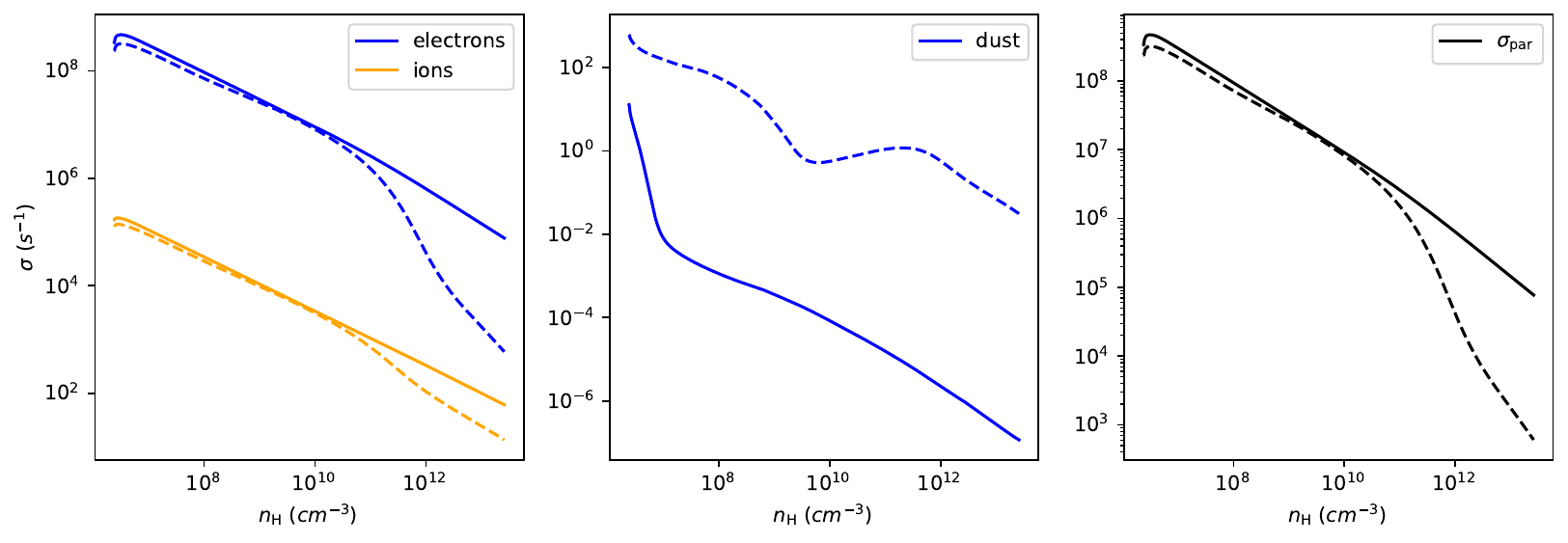}

    \caption{Individual conductivities profiles. The third panel displays the parallel conductivity. The solid lines correspond to the solid grain case ($\gamma_\mathrm{grain} = 190 \ \mathrm{erg \ cm^{-2}}$) and the dashed lines to the fragile grain case ($\gamma_\mathrm{grain} = 10 \ \mathrm{erg \ cm^{-2}}$).} 
    \label{par conductivity profile}
\end{figure*}

\begin{figure*}
    \includegraphics[width= 1.0\textwidth]{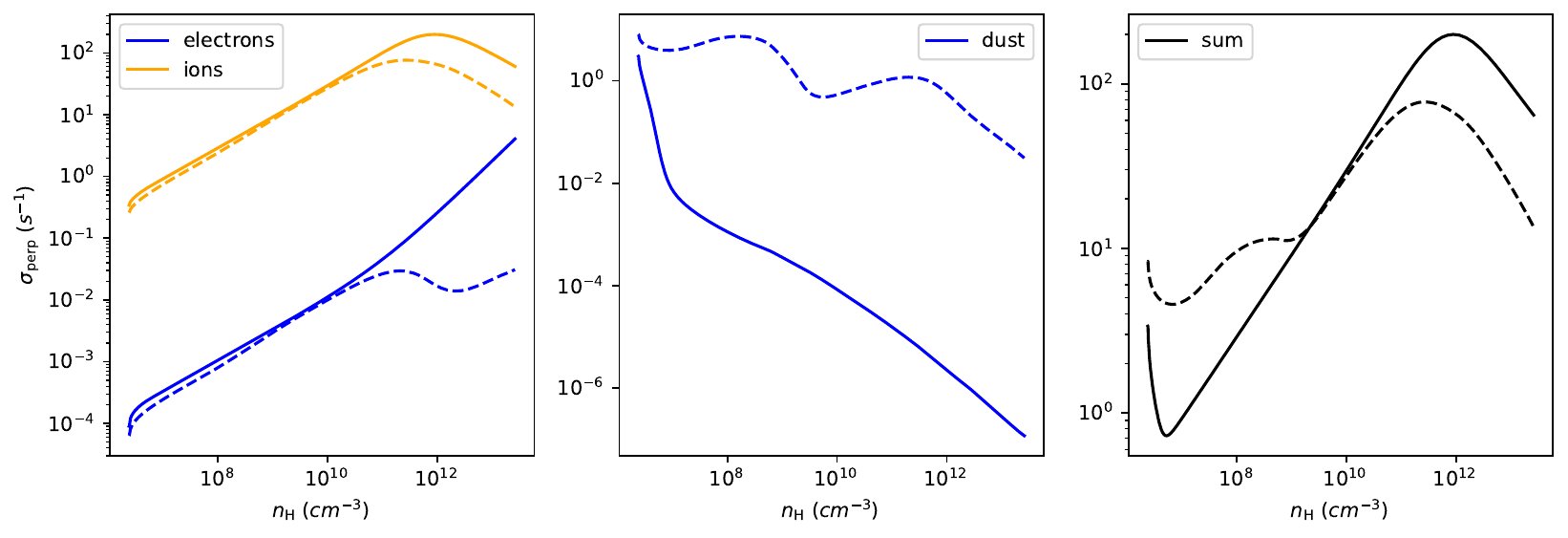}

    \caption{Perpendicular individual conductivities profiles. The third panel displays the total perpendicular conductivity, taken as the sum of the different contributions. The solid lines correspond to the solid grain case ($\gamma_\mathrm{grain} = 190 \ \mathrm{erg \ cm^{-2}}$) and the dashed lines to the fragile grain case ($\gamma_\mathrm{grain} = 10 \ \mathrm{erg \ cm^{-2}}$).} 
    \label{perp conductivity profile}
\end{figure*}

\begin{figure*}
    \includegraphics[width= 1.0\textwidth]{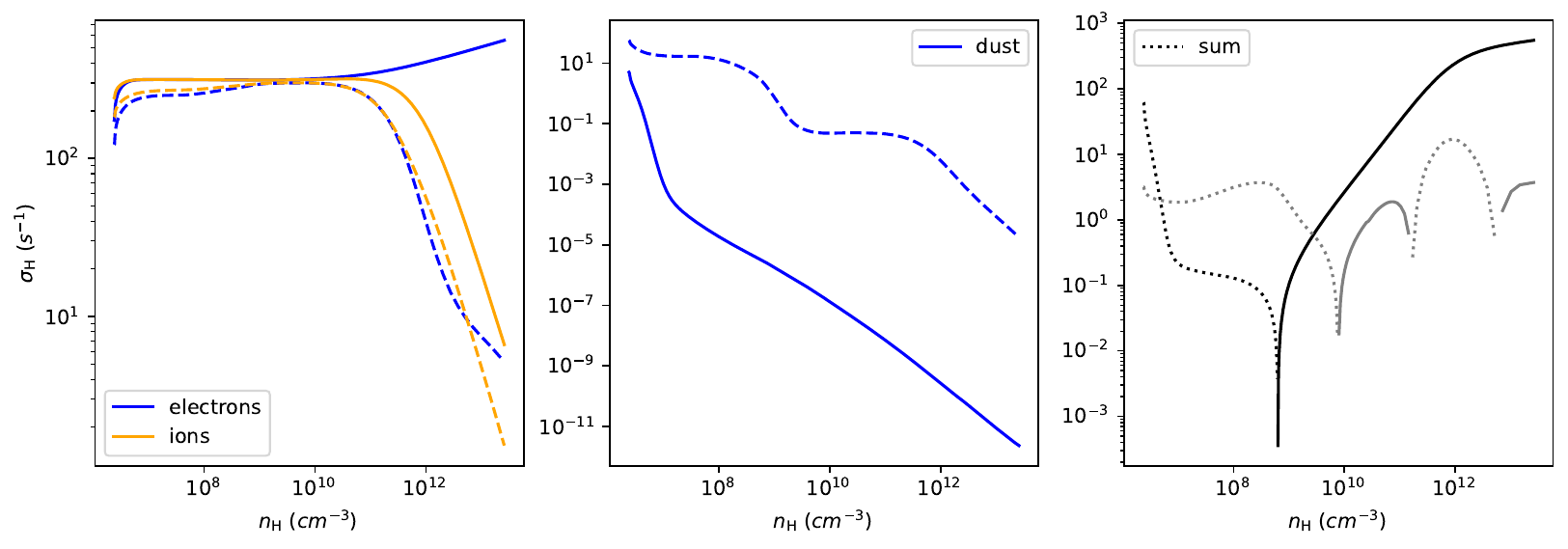}

    \caption{Hall individual conductivities profiles. The solid lines correspond to the solid grain case ($\gamma_\mathrm{grain} = 190 \ \mathrm{erg \ cm^{-2}}$) and the dashed lines to the fragile grain case ($\gamma_\mathrm{grain} = 10 \ \mathrm{erg \ cm^{-2}}$). The third panel displays the total Hall conductivity, taken as the sum of the different contributions, where the dotted parts are the negative values and the solid parts the positive ones. The solid grain case appears in black while the fragile case appears in grey.} 
    \label{Hall conductivity profile}
\end{figure*}
Figure \ref{par conductivity profile} explicitly shows that the individual electronic conductivity largely prevails over the others by a few orders of magnitude. Consequently, the total parallel conductivity is controlled by the electrons. In reducing the surface energy of the grains, $\gamma_\mathrm{grain}$, we allow for the small grains to be repopulated and to efficiently collect the electrons, resulting in a weaker overall parallel conductivity. Since the collision rate between small grains and electrons increases with density (regardless of dust coagulation), the impact on the electron conductivity and thus on the parallel conductivity only appears tardily in the collapse. Indeed, the dashed line diverges from the solid line at high densities only. The same reasoning is valid for the ions whose conductivity decreases as they recombine with electrons at the grains surface.

Regarding the individual perpendicular conductivities, the electronic contribution is negligible. Indeed, Fig. \ref{perp conductivity profile} suggests that the total perpendicular conductivity is dominated by both the dust and the ions at low density. Eventually as the density goes up, the dust influence vanishes and the ions prevail. In enhancing fragmentation, the replenishment of small grains gives rise to a higher dust perpendicular conductivity (the large grains contribution is insignificant) which leads to a higher total perpendicular conductivity at low  density ($n_\mathrm{H} \leq 10^{9} \ \mathrm{cm^{-3}}$). This effect is instantaneous since it does not involve any capture or interaction whatsoever. On the third panel, we can see that the dashed and solid lines overlap as soon as the dust ceases to be the dominant species, and the detachment between both curves at high densities is due to the ion density diminution (caused by electron/ion recombination) being significant at high densities only. We can derive an analytical formulae of a threshold value above which the ion density allows for the ion perpendicular conductivity to be larger than the dust perpendicular conductivity, namely, when $\sigma_\mathrm{perp,ion} > \sigma_\mathrm{perp,dust}$. When this condition is satisfied, the ambipolar resistivity $\eta_\mathrm{AD}$ becomes independent on the number of small grains, which explains why strong variations in the population of small grains sometimes do not reflect on the ambipolar resistivity (see for example Fig. \ref{res_erosion} at $n_\mathrm{H} = 10^8 \ \mathrm{cm^{-3}}$). Using the expressions of Eq. (\ref{conductivities expressions}), we get:

\begin{equation}
\int_{s_\mathrm{min}}^{s_\mathrm{max}}\frac{dN_\mathrm{d}}{ds}\frac{m_\mathrm{d}}{\tau_\mathrm{d}}ds < \frac{n_\mathrm{ion} \mu_\mathrm{ion} m_\mathrm{H}}{\tau_\mathrm{ion}},  
\end{equation}
where $m_\mathrm{d}$, $\tau_\mathrm{d}$, $n_\mathrm{ion}$, $\mu_\mathrm{ion}$ and $\tau_\mathrm{ion}$ are respectively the dust mass and reduced temperature, the ion density, mean ion molecular weight and reduced temperature. $\frac{dN_\mathrm{d}}{ds} = As^{-q}$ is the dust number distribution taken as a power law of index $q$ and $A = \frac{(4-q)\rho_\mathrm{dust}}{\frac{4}{3 \pi}\rho_\mathrm{grain}\mu_\mathrm{gas} |s_\mathrm{max}^{4-q}-s_\mathrm{min}^{4-q}|}$ is a normalisation factor
with $s_\mathrm{max}$ and $s_\mathrm{min}$ being the uppermost and lowermost sizes of the dust distribution considered. Integrating for any $q < 4$, we get:

\begin{equation}
    n_\mathrm{ion} > \left\{
    \begin{array}{ll}
       \frac{K \left(n_\mathrm{H},T \right)}{s_\mathrm{max}} log \left (\frac{s_\mathrm{max}}{s_\mathrm{min}} \right ), \mbox{if q = 3}  \\
      \frac{K \left(n_\mathrm{H},T \right)}{s_\mathrm{max}^{4-q}} \frac{4-q}{3-q} \left [s_\mathrm{max}^{-q+3}-s_\mathrm{min}^{-q+3} \right ], \mbox{otherwise}.
 
        \end{array}
        \right.
\end{equation}
Here, $K \left(n_\mathrm{H},T \right) = \frac{\epsilon m_\mathrm{H} \pi^{2} \mu_\mathrm{gas} n_\mathrm{H} c_\mathrm{s} \tau_\mathrm{ion}}{\rho_\mathrm{grain} \mu_\mathrm{ion} \sqrt{\frac{\pi \gamma}{8}}}$ is a factor which depends on the gas density and the temperature. Considering  $\frac{s_\mathrm{min}}{s_\mathrm{max}} \ll 1$, we can simplify the expressions to: 

\begin{equation} \label{ion density threshold}
    n_\mathrm{ion} > \left\{
    \begin{array}{ll}
       \frac{K \left(n_\mathrm{H},T \right)}{s_\mathrm{max}}  \left (\frac{4-q}{3-q} \right ), \mbox{if \ } q < 3  \\
      \frac{K \left(n_\mathrm{H},T \right)}{s_\mathrm{max}} \left|\frac{4-q}{3-q}\right| \left (\frac{s_\mathrm{max}}{s_\mathrm{min}} \right)^{|3-q|}, \mbox{if \ } q > 3.
 
        \end{array}
        \right.
\end{equation}
We can see that in reducing $s_\mathrm{min}$, the dust contribution would be more important and, thus, the ion density would have to be higher in order to maintain the dominance of the ion contribution. In increasing $s_\mathrm{max}$, for a given $q$ and $s_\mathrm{min}$, we reduce the number of small grains and consequently relax the constrain on $n_\mathrm{ion}$ which can afford to be lower. We stress that to apply this equation at some density $n_\mathrm{H}$, one has to describe precisely the dust distribution, that is to know $s_\mathrm{max}$, $s_\mathrm{min}$ and $q$. To gain insight, we apply Eq. (\ref{ion density threshold}) at the initial conditions of the collapse for which the shape of the distribution is known, since it is set as an MRN-like power law. With $s_\mathrm{min} = 5 \ \mathrm{nm}$, $s_\mathrm{max} = 250 \ \mathrm{nm}$, $q = 3.5$, $n_\mathrm{H} = 2.6 \times 10^{6} \ \mathrm{cm^{-3}}$ and $T = 10 \ \mathrm{K}$, we get $n_\mathrm{ion} > 1.67 \ \mathrm{cm^{-3}}$, and we can see on the various plots that initially $n_\mathrm{ion} \thicksim 10^{-2} \ \mathrm{cm^{-3}}$, which means it is indeed the dust contribution which prevails in regard to the perpendicular conductivity at that particular moment.

When it comes to the Hall conductivity, the individual ion conductivity yields negative values while those of the electrons and the dust yield positive values. Thus, there is a balance of power between the two groups in regard to the sign of the total Hall conductivity. At low densities, the electronic and ionic values are very close to each other, the ion conductivity being slightly superior. The dust contribution is only one order of magnitude lower but yet insufficient to reverse the trend directly. However, at some point during the collapse, the electron curve gets over the ion curve, resulting in the total Hall conductivity switching from negative to positive sign. Upon comparison with the ion conductivity, the electron conductivity is more strongly influenced by the population of small grains and since both conductivities are very close to each other throughout the collapse, the least variation in the population of small grains can affect the electronic capture and thus induce a change of sign of the Hall conductivity. That is why the Hall conductivity and thus the Hall resistivity are very sensitive to the overall dust distribution, large grains included. For instance, if the fragmentation threshold is shifted towards higher sizes, more massive grains can survive collisional events, and as a consequence the mass available for the repleshishment of small grains is lower.
On the first panel, when fragmentation is curtailed (solid line), the small grains are significantly depleted, allowing the electron conductivity to dwell on large values. Initially, the ion conductivity is above the electronic one but gets beneath it and ends up sinking, yielding an early transition in the total Hall conductivity sign, slightly below $n_\mathrm{H} = 10^{9} \ \mathrm{cm^{-3}}$. Nevertheless when considering fragile grains (dashed line), the early fragmentation triggers a significant replenishment of the small grains, resulting in a electronic conductivity decreasing and remaining below the ion conductivity until $n_\mathrm{H} = 10^{10} \ \mathrm{cm^{-3}}$ where the electrons take over. That is why the first transition occurs slightly later. Thereafter, the ion contribution recovers dominance slightly prior to $n_\mathrm{H} = 10^{12} \ \mathrm{cm^{-3}}$ to finally let the electron contribution prevail again at the end of the collapse. In this situation, the transition occurs three times during the collapse.

\section{Comparison between fixed MRN and coagulating distribution with and without fragmentation}

\begin{figure*}[hbt]
    \includegraphics[width= 1.0\textwidth]{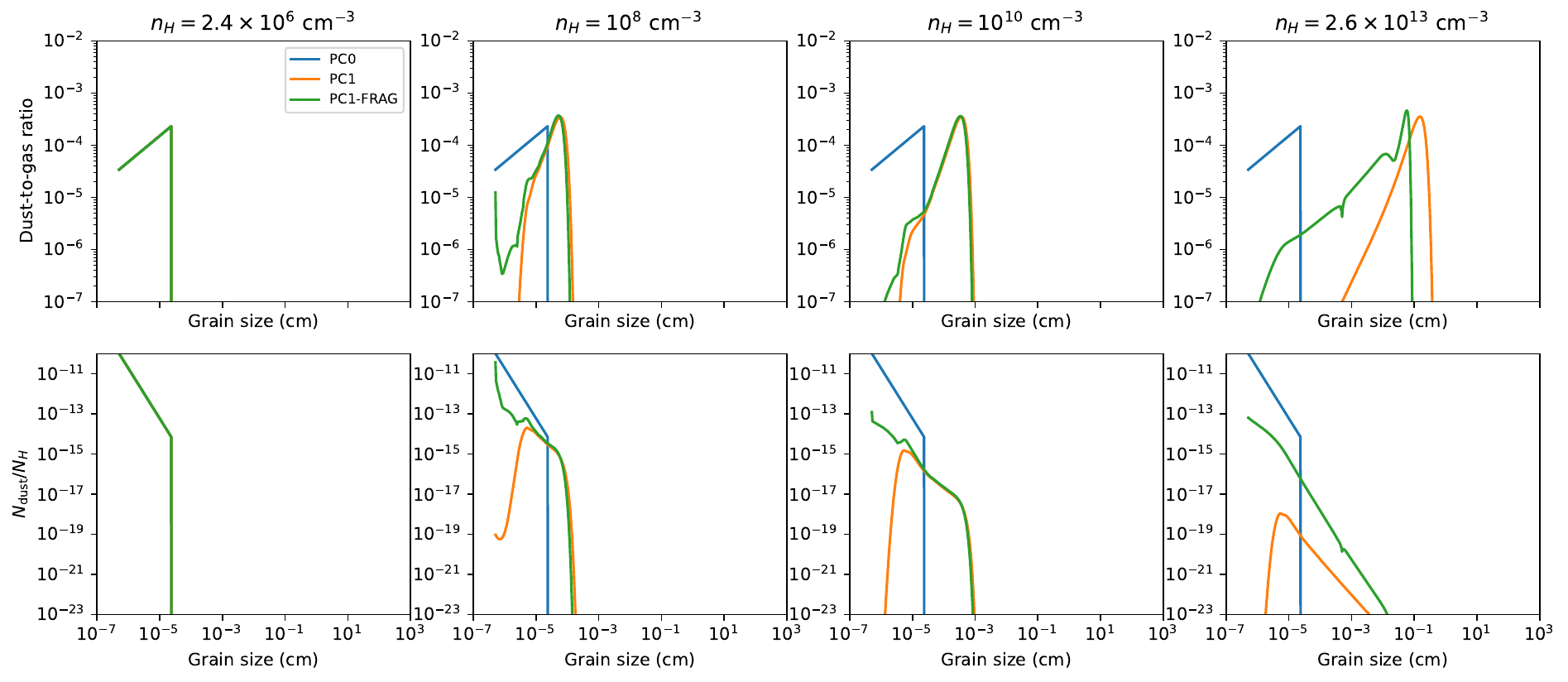}

    \caption{Dust distributions at different stages during the collapse. Comparison between the MRN non-evolving dust ($\mathrm{PC0}$), the coagulation only ($\mathrm{PC1}$), and the coagulation-fragmentaion ($\mathrm{PC1-FRAG}$) models. Top panels: Mass dust-to-gas ratio. Bottom panels: Number dust-to-gas ratio. } 
 
\end{figure*}

\begin{figure*}[hbt]
    \includegraphics[width= 1.0\textwidth]{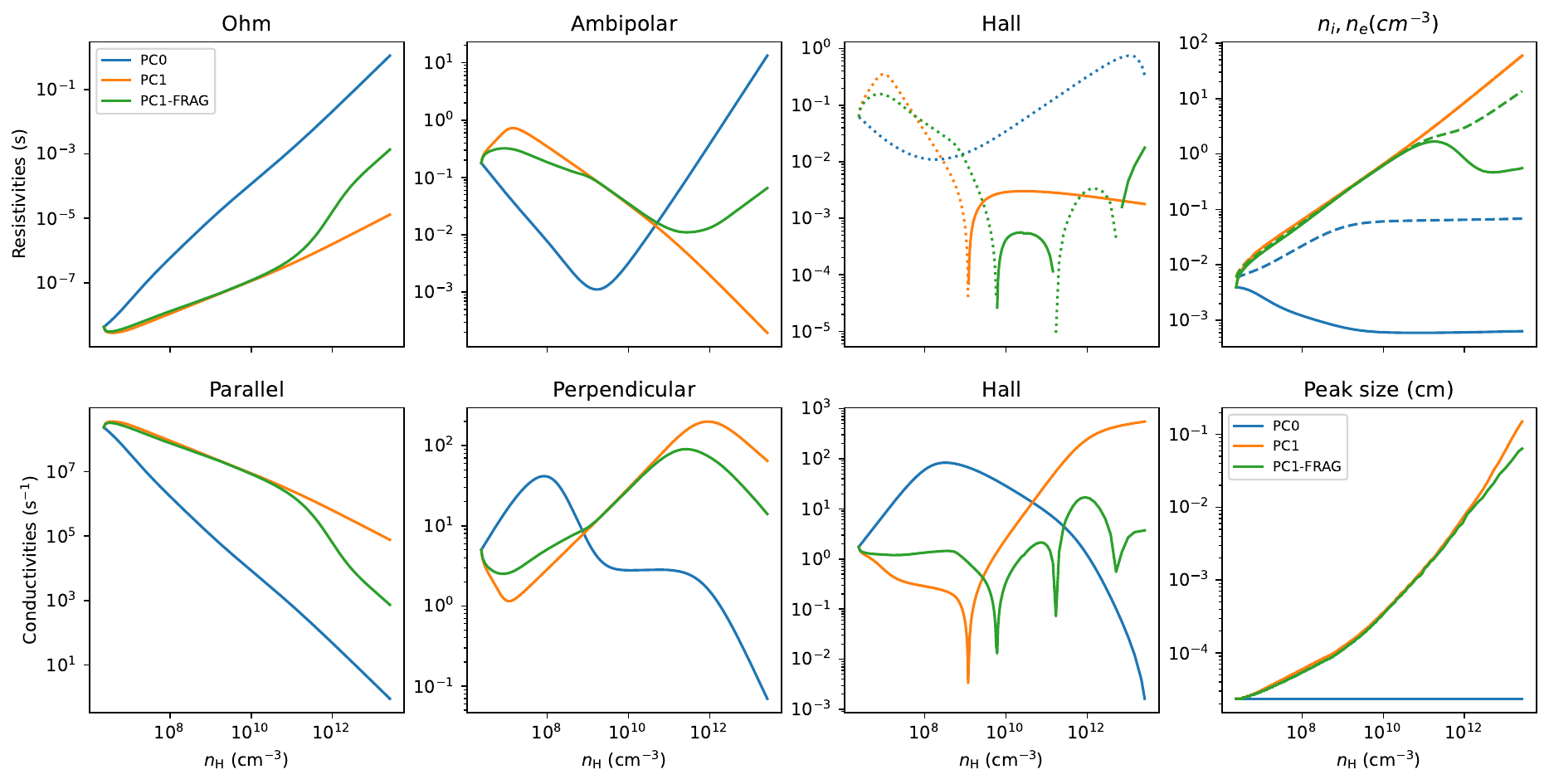}

    \caption{Conductivities and resistivities at different stages during the collapse (Hall resistivity : dotted line for the negative values and solid line for the positive ones). Comparison between the MRN non-evolving ($\mathrm{PC0}$), the coagulation only ($\mathrm{PC1}$), and the coagulation-fragmentaion ($\mathrm{PC1-FRAG}$) dust models. Last column: Electrons (solid) and ions (dashed) density (top), and peak size of the dust distribution (bottom).} 
    \label{coag_frag_comparison}
\end{figure*}

In Fig. \ref{coag_frag_comparison}, we offer a  simple comparison of the coagulation and fragmentation algorithms with other works. We use the same nomenclature as in \cite{Lebreuilly2023} to facilitate comparison. $\mathrm{PC0}$ refers to the MRN non-evolving dust model, $\mathrm{PC1}$ to the coagulating dust model and $\mathrm{PC1-FRAG}$ to the coagulating plus fragmenting dust model. $\mathrm{PC}$ stands for protostellar collapse. It shows results similar to those in \cite{Lebreuilly2023} and \cite{Kawasaki2022}. Note that for the sake of relevancy, in this peculiar case, we used the same value for the surface energy as the authors, namely, $\gamma_\mathrm{bare,si} = 25 \  \mathrm{erg \ cm^{-2}}$, along with the appropriate reduced elastic modulus value for bare-silicate grains. In particular, we can clearly see the replenishment of small grains in $\mathrm{PC1-FRAG}$ as compared with $\mathrm{PC1}$, when fragmentation is included. This leads to a depletion of electrons at the latest stages of the collapse, and consequently to a change in the slopes of the ohmic and ambipolar resistivities.

\section{Considering the most efficient source of relative velocity for the removal of small grains }

\begin{figure*}[hbt]
    \includegraphics[width= 1.0\textwidth]{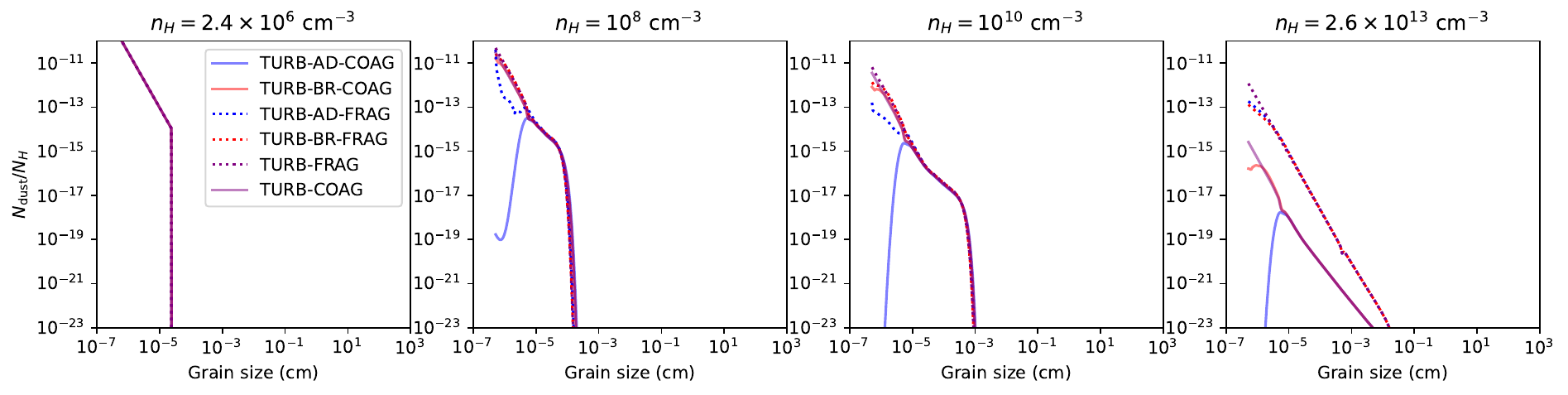}

    \caption{Dust number distributions at different stages during the collapse. Comparison between different sources of relative velocities with and without fragmentation : $\mathrm{TURB}$ (turbulence only), $\mathrm{TURB-AD}$ (turbulence plus ambipolar diffusion) and $\mathrm{TURB-BR}$ (turbulence plus Brownian motion). The solid lines refer to the purely coagulating cases, while fragmentation is included for the dashed lines ($\gamma_\mathrm{grain} = 10 \ \mathrm{erg \ cm^{-2}}$).}  
    \label{velocity_sources_comparison}
\end{figure*}

\label{comparison_sources}

\begin{figure*}[hbt]
    \includegraphics[width= 1.0\textwidth]{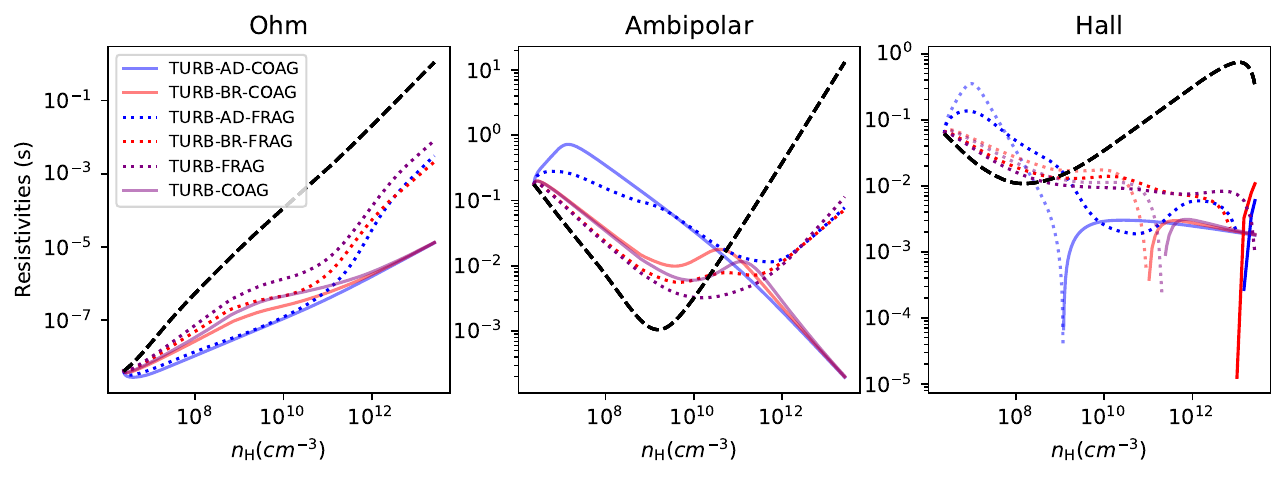}

    \caption{Magnetic resistivities at different stages during the collapse. Comparison between different sources of relative velocities with and without fragmentation : $\mathrm{TURB}$ (turbulence only), $\mathrm{TURB-AD}$ (turbulence plus ambipolar diffusion) and $\mathrm{TURB-BR}$ (turbulence plus brownian motion). The solid lines refer to the purely coagulating cases, while fragmentation is included for the dashed lines. The dashed black lines represent the non-evolving MRN dust distribution case.}  
    \label{velocity_sources_comparison_res}
\end{figure*}

Figure \ref{velocity_sources_comparison} provides a comparison of the effect of the different sources of grain collisions on the removal of small grains. $\mathrm{TURB-COAG}$ model includes only turbulence as a source of grain collisions. $\mathrm{TURB-BR-COAG}$ includes additionally Brownian motion while $\mathrm{TURB-AD-COAG}$ contains ambipolar diffusion. The models displaying $\mathrm{FRAG}$ account for fragmentation in addition to coagulation. We see that in the case of pure coagulation, the number of small grains is much lower when ambipolar diffusion is included, while it remains higher when only brownian motion is accounted for, at all densities. Therefore, in this context,  ambipolar diffusion is the main process in regard to the depletion of small grains.
However when including fragmentation, it seems that the gap in small grain removal efficiency between ambipolar diffusion and brownian motion is clearly reduced as fragmentation allows for a systematic replenishment of the small grains. At intermediate densities, ambipolar diffusion still appears to be slightly more efficient at removing the small grains, whereas at high density (last column), both velocity sources are contributing  to the depletion of small grains with Brownian motion being slightly more efficient. Note that since $\mathrm{TURB-AD-FRAG}$, $\mathrm{TURB-BR-FRAG,}$ and $\mathrm{TURB-FRAG}$ curves are not completely overlapping, small grain depletion is not completely suppressed by the repopulation induced by fragmentation. 
In addition, although there exists a striking difference in small grain population between models that include ambipolar diffusion and those which do not, we still observe a reduction of a few orders of magnitude in the overall number of small grains during the collapse (even more striking when fragmentation is off) regardless of the velocity source at play. Consequently, we can see on Fig. \ref{velocity_sources_comparison_res} that even without ambipolar drift, the ambipolar resistivity, $\eta_\mathrm{AD}$, reaches values larger than the MRN case.

\section{Ambipolar drift intensity in a collapsing core}
\label{delta in PPD}

\begin{figure*}[h!]
\centering
\begin{subfigure}{0.49\textwidth}
    \includegraphics[width=\textwidth]{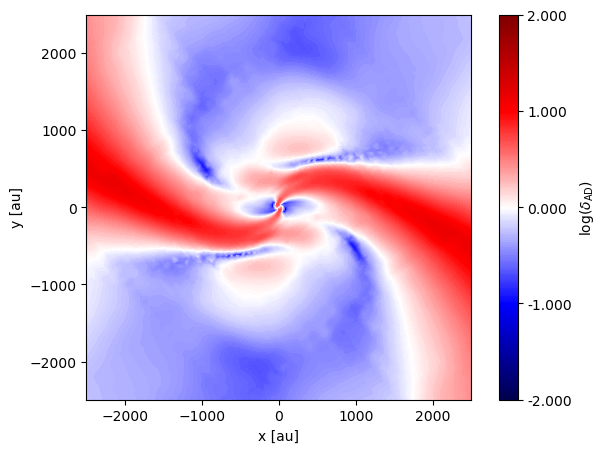}
    \caption{Face-on view.}
    \label{face on}
\end{subfigure}
\hfill
\begin{subfigure}{0.49\textwidth}
    \includegraphics[width=\textwidth]{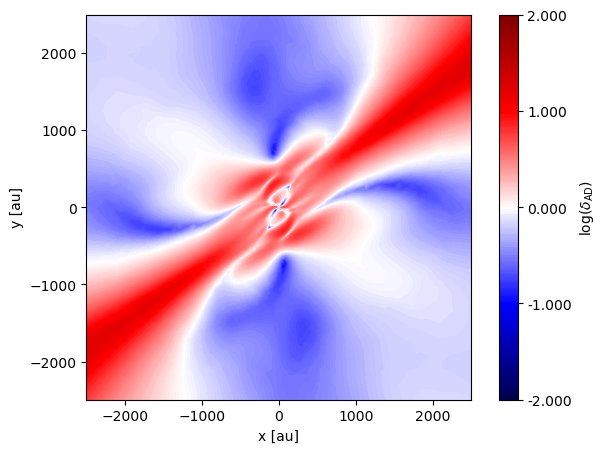}
    \caption{Edge-on view.}
    \label{edge on}
\end{subfigure}

\caption{Face-on (left) and edge-on (right) slices of the ambipolar drift intensity, $\delta_\mathrm{AD}$, extracted from a non-ideal MHD collapse model \citep[mmMRNmhd, ][]{Lebreuilly2020} computed with RAMSES \citep{Teyssier2002} at $t = 80 \ \mathrm{Kyr}$.  }
\label{PPD_plots}
\end{figure*}

Here is a map of the ambipolar drift intensity $\delta_\mathrm{AD}$ in a collapsing core extracted from a 3D \texttt{RAMSES} \citep{Teyssier2002} simulation performed by \cite{Lebreuilly2020}, with the  mmMRNmhd model, at $t = 80 \ \mathrm{Kyr}$ (Fig. \ref{delta in PPD}). It is inferred by measuring the Lorentz force $\left(\vec{\nabla} \times \vec{B} \right) \times \vec{B}$ and comparing it with the approximation $\frac{|\vec{B}|^2}{\lambda_\mathrm{J}}$ (see Eq. \ref{AD vectorial form}). As clearly seen, the intermediate value $\delta_\mathrm{AD} = 1$ is scarce, while both larger and lower values are ubiquitous. In the most central region (very young, nascent protoplanetary disc and core), and at large scale in the pseudo-disc, the ambipolar drift intensity dwells around $\delta_\mathrm{AD} = 10$. Consequently, grain-grain erosion is expected to induce a significant replenishment of the population of small grains in those denser regions. In contrast, the ambipolar drift intensity is much weaker ($\delta_\mathrm{AD} \sim 0.1$) in the rest of the domain, which hints at an inhibition of small grain coagulation via electrostatic repulsion in those diffuse regions. This reinforces the fundamental role that both mechanisms, electrosatic repulsion and grain-grain erosion, play in maintaining a reasonable ambipolar resistivity, $\eta_\mathrm{AD}$, during the protostellar collapse.

\section{Impact of grain-grain erosion and electrostatic repulsion on the Ohm resistivity}

\begin{figure*}[hbt]
    \centering
    \includegraphics[width= 0.8\textwidth]{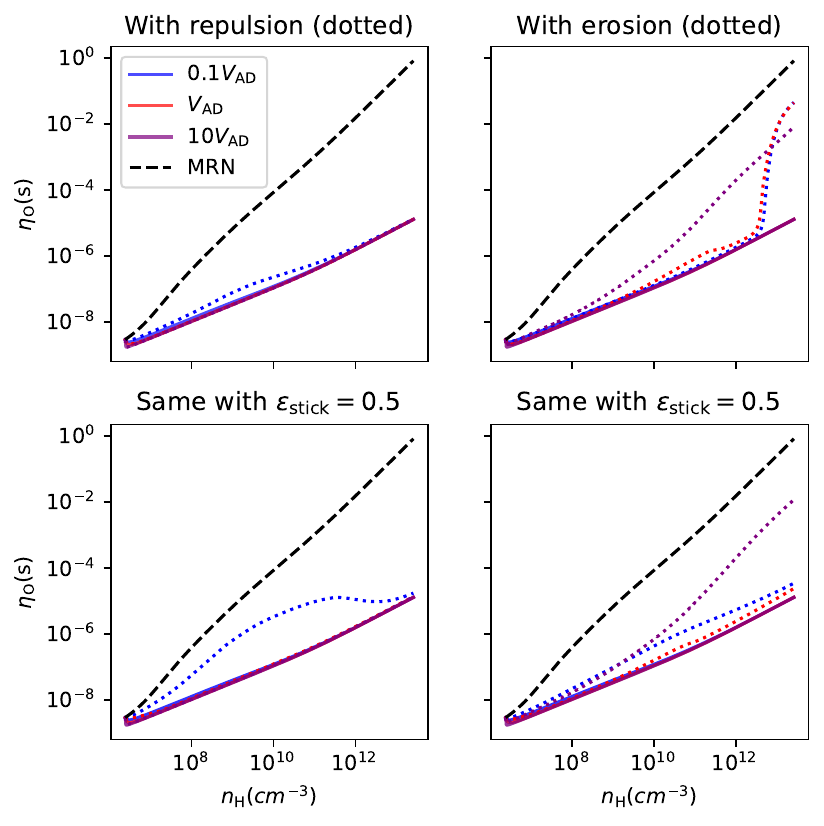}
    \caption{Ohm resistivity, $\eta_\mathrm{O}$, evolution during the protostellar collapse for different values of the ambipolar drift intensity or equivalently different ambipolar drift velocities: $\delta_\mathrm{AD} = 0.1 \  (0.1V_\mathrm{AD})$, $\delta_\mathrm{AD} = 1 \ (V_\mathrm{AD})$, and $\delta_\mathrm{AD} = 10 \ (10V_\mathrm{AD})$.  Grain-grain electrostatic repulsion (first column) and erosion (second column) are also included (dotted lines). Second row:\ Same, but with a sticking efficiency of 0.5.}
    \label{resOHM_electrostatic_repulsion+delta}
\end{figure*}

Similarly to the analysis carried out in Sect. \ref{result impact of amibpolar drift}, we investigated the impact of grain-grain electrostatic repulsion and erosion on the Ohm resistivity $\eta_\mathrm{0}$ for different values of the ambipolar drift intensity $\delta_\mathrm{AD}$. Looking at the first column, we can see that contrary to the ambipolar resistivity, the Ohm resistivity remains almost unaffected at all densities even for a weak ambipolar drift intensity  ($\delta_\mathrm{AD} = 0.1$) for which electrostatic repulsion is expected to significantly inhibit small grains coagulation. While the dust contributes directly to the ambipolar resistivity, $\eta_\mathrm{AD}$, at low density (any change in the small grain population is reflected almost immediately in the resistivity profile), it only acts indirectly on the ohmic resistivity, via electron capture. At low densities, the cross-section for electron capture is very low and only a sufficient abundance of small grains, such as the one enabled by a lower sticking efficiency ($\epsilon_\mathrm{stick} = 0.5$, bottom left panel), allows for an efficient collection of electrons and, thus, a rise in the ohmic resistivity. At high density, the grain collisional velocities are usually above the Coulomb velocity threshold (Eq. \ref{Coulomb barrier}), thereby suppressing the effect of electrostatic repulsion.

When it comes to the impact of erosion (right column), it is clear that for low and intermediate values of ambipolar drift intensity ($\delta_\mathrm{AD} = 0.1$ and $\delta_\mathrm{AD} = 1$), the collisional velocities are not high enough to induce a sufficient replenishment of small grains prior to $n_\mathrm{H} = 10^{12} \ \mathrm{cm^{-3}}$. However, beyond this density, the collisional velocities and dust size ratios $s_\mathrm{ratio}$ (see Eq. \ref{erosion efficiency equation}) are large enough to enhance the erosion efficiency and thus to allow for the excess of small grains to efficiently capture the electrons and to allow for a soaring in terms of the Ohm resistivity. Nevertheless, in order to increase the resistivity at lower density, we need to rely on a stronger ambipolar drift intensity $\delta_\mathrm{AD} = 10$, as seen by looking at the purple dotted curve. We note that for $\delta_\mathrm{AD} = 0.1$ and $\delta_\mathrm{AD} = 1$, a sticking efficiency of $\epsilon_\mathrm{stick} = 0.5$ does reduce the small grain coagulation rate, but is also detrimential to erosion, since only half of the collisions are destructive. This results in an overall suppression of the late rise in ohmic resistivity.

In any case, the Ohm resistivity remains at least two orders of magnitude below the MRN reference because no process is capable of completely making up for the very short coagulation timescales. Besides, it remains dominated by the ambipolar resistivity $\eta_\mathrm{AD}$ at least up to the final density value of our simulations, namely, $n_\mathrm{H} = 2 \times 10^{13} \ \mathrm{cm^{-3}}.$

\section{Impact of the grain-grain sticking efficiency, $\epsilon_\mathrm{stick}$}

\begin{figure*}[hbt]

    \includegraphics[width= 1.0\textwidth]{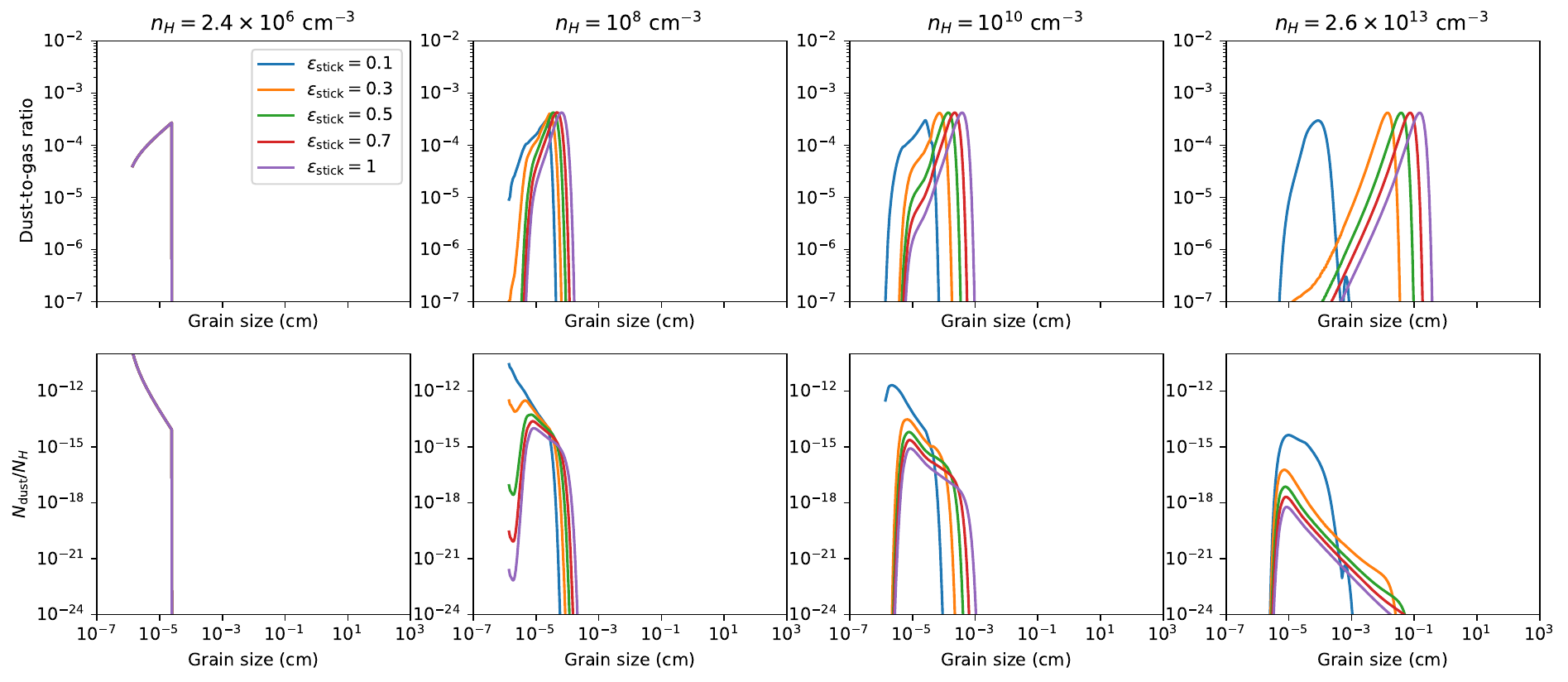}

    \includegraphics[width= 1.0\textwidth]{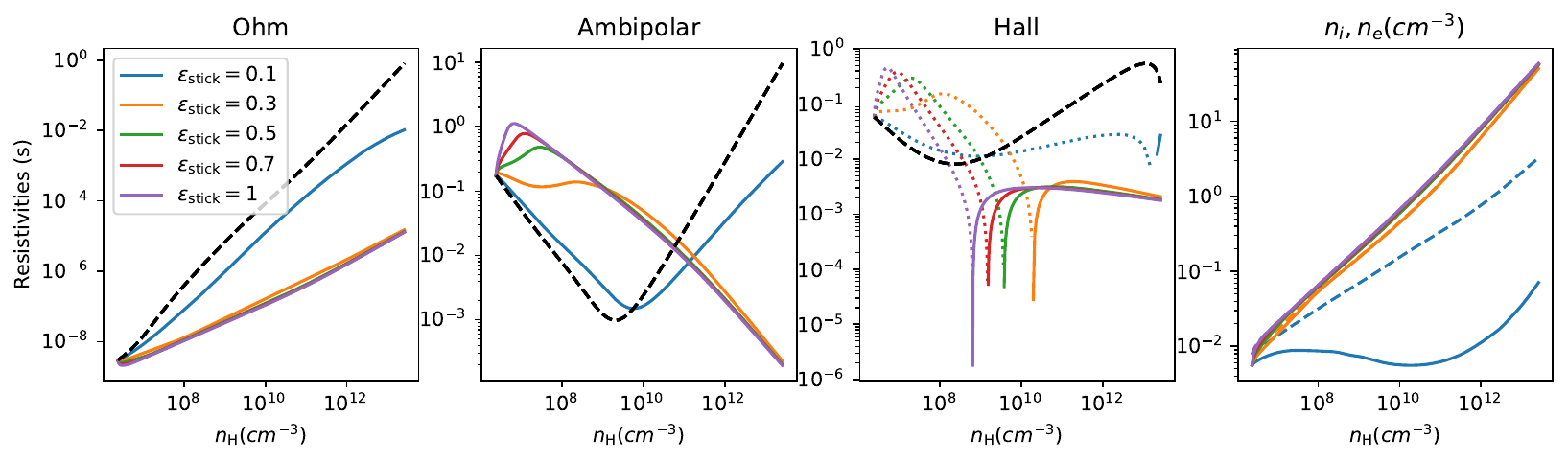}

    \caption{Same as Fig. \ref{res_gamma_comparison_icenopermfrag}, but investigating the grain sticking efficiency ($\gamma_\mathrm{grain} = 190 \ \mathrm{erg} \ \mathrm{cm^{-2}}$).} 
    \label{res_sticking_efficiency}
\end{figure*}

Here, we vary the sticking efficiency of grains, $\epsilon_\mathrm{stick}$, (included in our coagulation algorithm) and explore its impact on the evolution of the dust distribution as well as on the magnetic resistivities. We note that $\epsilon_\mathrm{stick} = 0.5$ implies that half of the collisions result in bouncing rather than coagulation or fragmentation, namely, that both grains remain intact. Figure \ref{res_sticking_efficiency} indicates that a value of $0.1, $ which would significantly hinder the small grains coagulation at low density, allowing us to recover the MRN resistivities to some extent. We note that a value of 0.1 means that only one collision out of ten leads to grain sticking, thus leading to coagulation. A value of $0.3$ slows down the coagulation process (as readily seen on the ambipolar resistivity $\eta_\mathrm{AD}$ profile) and allows for the ambipolar resistivity to be notably reduced prior to $n_\mathrm{H} = 10^{9} \ \mathrm{cm^{-3}}$. However, for values beyond $\epsilon_\mathrm{stick} = 0.5$, no significant changes are induced.
This parameter also has a clear impact on the peak size of the dust distribution. Indeed, a lower sticking efficiency implies a longer coagulation timescale and, consequently, a reduced dust peak size.

\end{document}